\documentclass{article}
\usepackage{spconf,amsmath,amssymb,graphicx}
\usepackage{cite,booktabs,multirow}
\usepackage[table]{xcolor}
\usepackage[ruled,vlined]{algorithm2e}
\usepackage{hyperref}

\usepackage{amsmath,amssymb}
\usepackage{graphicx}
\usepackage[most]{tcolorbox}

\usepackage{algorithmic}

\usepackage{newfloat}
\usepackage{listings}

\definecolor{promptframe}{HTML}{6E8796}
\definecolor{promptback}{HTML}{F4F7F8}
\definecolor{prompttitle}{HTML}{E2EBEF}
\newtcblisting{vlmprompt}[1]{
  enhanced,
  breakable,
  listing only,
  colback=promptback,
  colframe=promptframe,
  colbacktitle=prompttitle,
  coltitle=black,
  title={#1},
  fonttitle=\bfseries\small,
  boxrule=0.6pt,
  arc=1.2mm,
  left=1.5mm,
  right=1.5mm,
  top=1.2mm,
  bottom=1.2mm,
  before skip=7pt,
  after skip=8pt,
  listing options={
    basicstyle=\ttfamily\scriptsize,
    breaklines=true,
    breakatwhitespace=true,
    columns=fullflexible,
    keepspaces=true,
    showstringspaces=false
  }
}

\title{AcousticAgent: A Floorplan-Centric Simulation Platform for Distributed Multi-Room Speech Processing}

\name{
    Zhiqi Ai\textsuperscript{1},
    Han Cheng\textsuperscript{1},
    Shiyi Mu\textsuperscript{1},
    Yongjin Zhou\textsuperscript{1,*},
    Shugong Xu\textsuperscript{2,*}\thanks{\textsuperscript{*}Corresponding authors}
}

\address{
\textsuperscript{1}Shanghai University, Shanghai, China \\
\textsuperscript{2}Xi'an Jiaotong-Liverpool University, Suzhou, China
}

\begin{document}
\ninept
\maketitle
\begin{abstract}
Existing acoustic simulation tools either generate room impulse responses (RIRs) efficiently in parameterized rooms or rely on preconstructed three-dimensional scenes, making it difficult to build large numbers of editable residential multi-room environments. We present \textbf{AcousticAgent}, a floorplan-centric acoustic simulation platform that converts structured floorplans into editable, acoustically connected multi-room scenes. AcousticAgent jointly models room geometry and connectivity, furniture, materials, sound sources, and microphone configurations. A scene-oriented Acoustic Materials Database maps semantic elements to acoustic parameters, while a floorplan-aware hybrid geometrical-acoustics solver generates multichannel RIRs and reverberant audio. The platform also supports directional and moving sources, distributed microphones, and VLM-assisted scene construction from floorplan images and text. Using AcousticAgent, we construct \textbf{FP-RIR}, a dataset covering 15,376 residential floorplans with diverse room topologies, propagation conditions, and source--receiver configurations. We validate its acoustic fidelity against real-world measurements and evaluate whole-home source localization, speech enhancement, and far-field speech processing under single-device and distributed microphone settings. The results demonstrate the effectiveness of floorplan-aware modeling for large-scale residential acoustic simulation and the benefits of distributed spatial processing in multi-room environments. AcousticAgent and FP-RIR provide scalable and configurable resources for developing and evaluating residential speech systems.
\end{abstract}
\section{Introduction}
\label{sec:intro}

Residential speech interfaces often operate under far-field and cross-room conditions, where sound sources and microphones may be far apart, obstructed by furniture, or separated by intervening walls. The received signal therefore varies with room topology, source--receiver geometry, inter-room connectivity, and the acoustic properties of surrounding surfaces~\cite{savioja2015overview,vorlander2008auralization}. This spatial variability makes it difficult for a single microphone to provide uniform coverage throughout a home. Distributed microphone systems alleviate this limitation by combining spatially complementary observations and have been studied for source localization, distributed beamforming, speech enhancement, and multi-device speech processing~\cite{bertrand2010danse,heusdens2012distributedmvdr,grinstein2023gnnssl,ciccarelli2022multidevice}.

Developing and evaluating such systems requires acoustic data spanning diverse residential layouts, material properties, and source--microphone configurations. Measured corpora preserve real propagation effects, but each recording campaign is tied to a finite set of physical spaces and device deployments; broader coverage requires additional deployment and data collection~\cite{ravanelli2015dirha,qin2020ffsvc,guan2021libriadhoc40}. Room impulse response (RIR) simulation provides a scalable means of varying these factors, but its practical scalability depends not only on propagation computation, but also on how efficiently acoustic scenes can be specified. Parameterized simulators often require room geometry, materials, and device positions to be explicitly defined, which becomes cumbersome when many irregular multi-room layouts must be instantiated. Mesh-based platforms can represent richer three-dimensional environments, but rely on preconstructed scene assets~\cite{diaz2021gpurir,scheibler2018pyroomacoustics,chen2022soundspaces,li2025sonicsim}. Large-scale residential simulation therefore calls for a scene representation that is both acoustically meaningful and easy to construct, edit, and vary.

Structured floorplans provide a compact representation of room geometry, semantics, and connectivity. They can be parsed from raster images or generated from structural constraints and natural-language descriptions without requiring complete three-dimensional assets~\cite{kalervo2019cubicasa,deitke2022procthor,abouagour2025resplan,
galanos2023architext,liu2026floorplanvlm}. To form a computational acoustic scene, however, they must be augmented with furniture, acoustic materials, source and microphone configurations, and models of within-room and cross-room propagation. Although floorplans are widely used for visual parsing, layout generation, and embodied scene construction, their use as the core representation for large-scale acoustic simulation remains underexplored.

To address this gap, we present \textbf{AcousticAgent}, a floorplan-centric platform for multi-room acoustic simulation. AcousticAgent converts structured floorplans into editable, acoustically connected scenes that jointly represent room geometry and connectivity, furniture, materials, sound sources, and microphone configurations. A scene-oriented Acoustic Materials Database maps semantic scene elements to acoustic properties, while a floorplan-aware hybrid geometrical-acoustics solver models within-room and cross-room sound propagation to generate multichannel RIRs and reverberant audio. The platform further supports directional and moving sources, microphone arrays, distributed microphones, and VLM-assisted scene construction from floorplan images and textual descriptions.

Using AcousticAgent, we construct \textbf{FP-RIR}, a large-scale dataset spanning 15,376 residential floorplans with diverse room topologies, within-room and cross-room propagation conditions, and flexible source--receiver and distributed-microphone layouts. We validate the simulated acoustic characteristics against measured RIRs and establish a distributed multi-room speech-processing benchmark covering whole-home source localization, speech enhancement, and far-field speech processing under single-device and distributed microphone settings.

Our main contributions are summarized as follows:
\begin{itemize}

\item \textbf{Floorplan-centric multi-room acoustic simulation.}
We introduce AcousticAgent, which converts structured floorplans into editable, acoustically connected multi-room scenes through semantic acoustic-material assignment and floorplan-aware geometrical-acoustics simulation.

\item \textbf{Large-scale FP-RIR dataset.}
We construct FP-RIR spanning 15,376 residential floorplans with diverse room topologies, within-room and cross-room propagation conditions, and flexible source--receiver and distributed-microphone layouts.

\item \textbf{Distributed multi-room speech-processing benchmark.}
We establish a benchmark for whole-home source localization, speech enhancement, and far-field speech processing under single-device and distributed microphone settings.

\end{itemize}

\section{Related Work}

\subsection{Acoustic Simulation and RIR Generation}

Indoor acoustic simulation generates room impulse responses (RIRs) or
reverberant audio by modeling sound propagation according to scene geometry,
material properties, and source--receiver configurations. Existing approaches
are commonly categorized into wave-based and geometrical-acoustics methods
~\cite{thompson2006review,savioja2015overview}. Wave-based methods, such as
finite-element and finite-difference methods, solve the acoustic wave equation
and accurately capture interference, diffraction, and room modes, but their
computational cost increases rapidly with scene size and frequency range
~\cite{thompson2006review}. Geometrical-acoustics methods approximate
propagation using geometric paths and provide efficient alternatives through
image-source methods, ray tracing, and path tracing
~\cite{allen1979image,savioja2015overview,vorlander2008auralization}.

Existing simulation tools mainly target either efficient parameterized room
simulation or detailed three-dimensional scene rendering. RIR-Generator and
gpuRIR provide image-source-based simulation for parameterized rooms, while
FRAM-RIR and Pyroomacoustics support efficient multichannel RIR generation and
microphone-array processing~\cite{habets2006rir,diaz2021gpurir,luo2024framrir,
scheibler2018pyroomacoustics}. For more general environments, Steam Audio,
SoundSpaces 2.0, and SonicSim enable geometry-based acoustic rendering over
three-dimensional scene representations
~\cite{steamAudio,chen2022soundspaces,li2025sonicsim}. These systems provide
powerful acoustic rendering capabilities, but constructing large numbers of
editable residential multi-room scenes remains challenging, particularly when
layouts, materials, and source--receiver configurations need to be varied
systematically.

\subsection{Floorplan Representation and Scene Construction}

Structured floorplans compactly encode architectural geometry, room semantics, and spatial connectivity. CubiCasa5K supports architectural-element parsing from annotated raster floorplans, while FloorplanVLM converts raster images into structured vector representations ~\cite{kalervo2019cubicasa,liu2026floorplanvlm}. ResPlan provides large-scale residential layouts with explicit semantic and connectivity information ~\cite{abouagour2025resplan}. ArchiText and ProcTHOR further explore layout generation and procedural residential scene construction ~\cite{galanos2023architext,deitke2022procthor}. While structured layouts are widely used in visual understanding and embodied environment construction, acoustic simulation additionally requires material properties, source and microphone configurations, and cross-room propagation modeling. Their use as a central representation for large-scale acoustic scene construction remains underexplored.

\begin{figure*}[t]
    \centering
    \includegraphics[width=\textwidth]{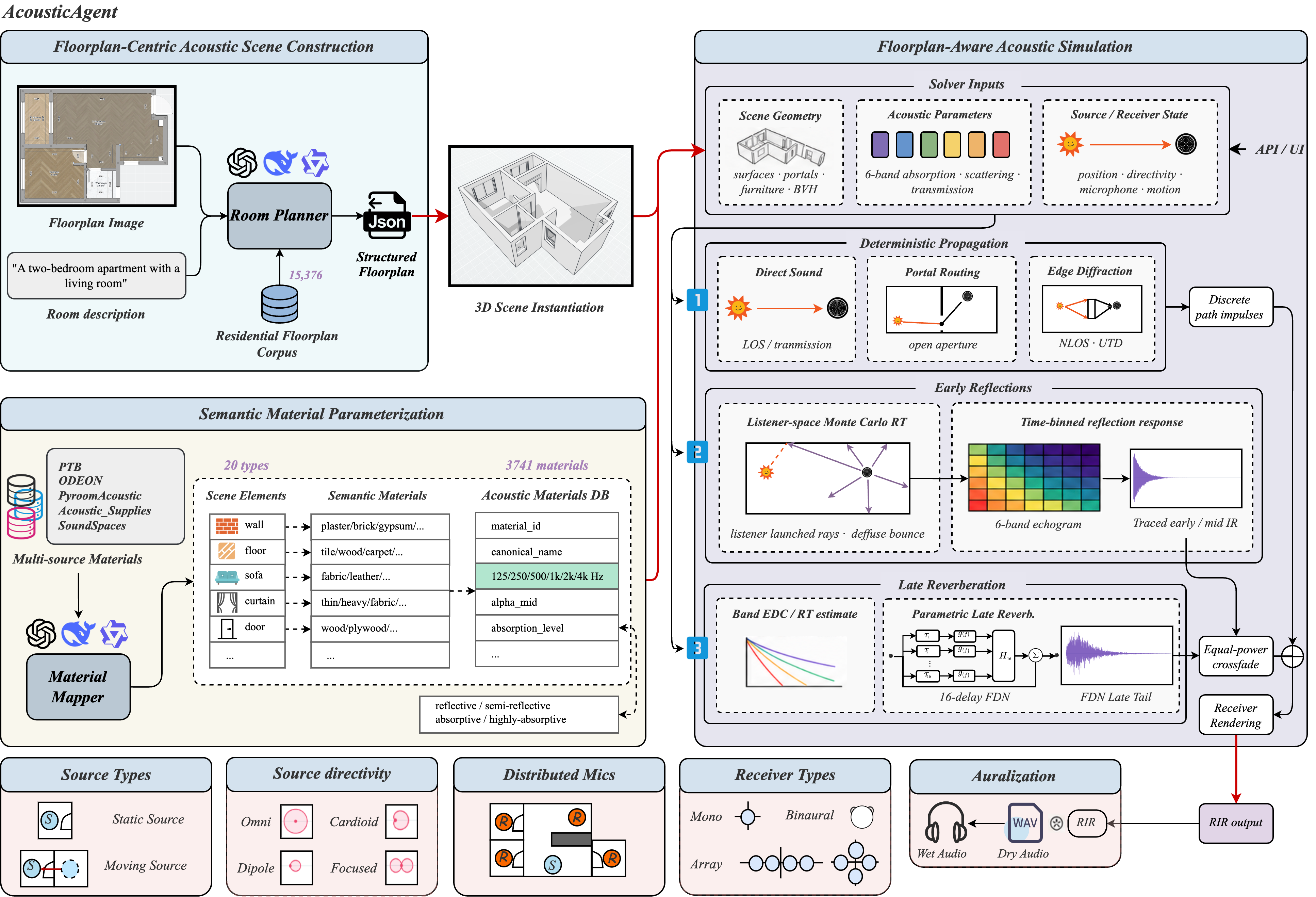}
    \caption{Overview of AcousticAgent.
    AcousticAgent transforms floorplan images, natural-language descriptions,
    or sampled residential layouts into editable multi-room acoustic scenes
    with explicit geometry, connectivity, furniture, and semantic material
    parameters. Its floorplan-aware hybrid solver combines deterministic
    propagation, Monte Carlo ray tracing, and parametric late reverberation to
    generate multichannel RIRs for static or moving sources and flexible
    microphone configurations.}
    \label{fig:acousticagent_overview}
\end{figure*}

\subsection{Distributed Multi-Room Speech Processing}

Distributed microphone systems exploit spatially separated observations for sound source localization, beamforming, speech enhancement, and speech separation~\cite{bertrand2010danse,heusdens2012distributedmvdr, grinstein2023gnnssl,wang2020distributed}. Recent work further combines multi-device acoustic front ends with downstream tasks such as keyword spotting, speaker verification, and automatic speech recognition ~\cite{yu2020multilookkws,liang2022multichannel,li2020multistream, ciccarelli2022multidevice}. Measured corpora are essential for evaluating real-world propagation and device effects. DIRHA-English provides distant-speech recordings in domestic environments, FFSVC 2020 targets far-field speaker verification with distributed microphone arrays, and Libri-adhoc40 contains synchronized recordings from 40 microphones in a real office ~\cite{ravanelli2015dirha,qin2020ffsvc,guan2021libriadhoc40}. However, such datasets cover only a limited number of environments, room topologies, and sensor layouts. Large-scale controllable simulation is therefore needed to complement real recordings and support systematic evaluation across diverse residential floorplans and microphone configurations.

\section{AcousticAgent}
\label{sec:acousticagent}

As shown in Fig.~\ref{fig:acousticagent_overview}, AcousticAgent transforms
residential floorplans into editable, acoustically connected multi-room scenes
for multichannel room impulse response (RIR) generation. Its key design is to
decouple reusable scene content---including geometry, connectivity, furniture,
and materials---from dynamic source and receiver configurations. Once a scene
has been constructed, it can be reused across different source positions,
trajectories, directivities, and microphone layouts. AcousticAgent comprises
four components: floorplan-centric acoustic scene construction, semantic
material parameterization, floorplan-aware acoustic simulation, and flexible
source and microphone configurations.

\subsection{Floorplan-Centric Acoustic Scene Construction}
\label{sec:scene_construction}

AcousticAgent accepts structured floorplans, raster floorplan images,
natural-language descriptions, or layouts sampled from a residential
floorplan corpus. Image- and text-based inputs may be processed with optional
LLM/VLM assistance, after which all inputs are normalized into the same
explicit and editable scene representation.

A structured floorplan is represented as
\begin{equation}
    \mathcal{P}
    =
    (\mathcal{R},\mathcal{W},\mathcal{O},\mathcal{F}),
    \label{eq:floorplan}
\end{equation}
where $\mathcal{R}$ denotes rooms, $\mathcal{W}$ denotes walls and other
boundary elements, $\mathcal{O}$ denotes doors, windows, and wall-free
openings, and $\mathcal{F}$ denotes furniture. Each room is associated with
its polygonal boundary, semantic type, and height. Interior openings connect
adjacent rooms and induce a room-connectivity graph
$\mathcal{G}_{\mathrm{room}}$, which explicitly represents both within-room
and cross-room relationships.

The two-dimensional layout is then instantiated as an editable
three-dimensional acoustic scene. Walls, ceilings, floors, and architectural
openings are constructed from the floorplan geometry, while furniture is
inserted as semantic geometry that may participate in occlusion and
reflection. This representation preserves the compactness and editability of
the original floorplan while providing the geometric structure required for
multi-room acoustic simulation.

\subsection{Semantic Material Parameterization}
\label{sec:material_parameterization}

To connect scene semantics with acoustic simulation, AcousticAgent introduces
a scene-oriented Acoustic Materials Database with 3,741 normalized records
across 20 scene-element categories. It consolidates octave-band acoustic data
from the PTB absorption database, the ODEON material library,
Pyroomacoustics, SoundSpaces 2.0, and Acoustic Supplies
~\cite{ptbAbsorptionDatabase,odeon2025manual,
scheibler2018pyroomacoustics,chen2022soundspaces,
acousticSuppliesAbsorption}. The database covers common residential elements,
including walls, floors, doors, glazing, curtains, and upholstered furniture.
Each record stores a canonical material name, provenance, octave-band
absorption coefficients, and auxiliary descriptors. Measured scattering and
transmission parameters are retained when available; otherwise, editable
category-level presets are used.

For scene element $e_i$ and frequency band $b$, its acoustic properties are
represented as
\begin{equation}
    \boldsymbol{\theta}_{i,b}
    =
    \left(
        \alpha_{i,b},
        \sigma_{i,b},
        \tau_{i,b}
    \right),
    \qquad
    b\in\mathcal{B},
    \label{eq:material_parameters}
\end{equation}
where $\alpha_{i,b}$, $\sigma_{i,b}$, and $\tau_{i,b}$ denote absorption,
scattering, and transmission coefficients, respectively, and
$\mathcal{B}=\{125,250,500,1000, \\2000,4000\}\,\mathrm{Hz}$.

Material candidates are constrained by scene-element semantics and may be
further refined through optional LLM/VLM-assisted matching. The selected
assignments are stored explicitly with the scene and remain manually editable,
allowing semantic scene descriptions to be converted into deterministic
acoustic parameters without coupling the solver to a specific recognition
model.

\subsection{Floorplan-Aware Acoustic Simulation}
\label{sec:acoustic_simulation}

Given the scene geometry, material parameters, and a source--receiver
configuration, AcousticAgent employs a hybrid geometrical-acoustics solver that
combines deterministic propagation, Monte Carlo ray tracing, and parametric
late reverberation. The room-connectivity graph guides the search for
cross-room routes, while the three-dimensional scene is used to validate
visibility, occlusion, and geometric interactions.

Deterministic propagation models perceptually important arrivals, including
line-of-sight direct sound, transmission through surfaces, propagation through
doors and wall-free openings, and edge-diffraction paths in non-line-of-sight
conditions. Portal sequences connecting the source and receiver rooms are
enumerated on $\mathcal{G}_{\mathrm{room}}$ and validated against the scene
geometry. Diffraction is approximated using a UTD-inspired formulation
~\cite{kouyoumjian1974utd}.

Early and mid reflections are modeled using receiver-centered Monte Carlo ray
tracing. At each surface interaction, frequency-dependent absorption and
transmission reduce the propagated energy, while scattering controls its
specular and diffuse components. The resulting arrivals are accumulated into
bandwise echograms. Late reverberation is synthesized using a feedback delay
network whose decay is conditioned on band-dependent reverberation estimates
obtained through Schroeder backward integration
~\cite{schroeder1965reverberation,jot1992analysis}.

For receiver $q$, source state $k$, and frequency band $b$, the hybrid
response is written as
\begin{equation}
\begin{split}
    h_{q,k,b}[n]
    ={}&
    h^{\mathrm{det}}_{q,k,b}[n]
    +
    w_{\mathrm{rt}}[n]h^{\mathrm{rt}}_{q,k,b}[n]
    \\
    &+
    w_{\mathrm{late}}[n]h^{\mathrm{late}}_{q,k,b}[n],
\end{split}
    \label{eq:hybrid_rir}
\end{equation}
where $h^{\mathrm{det}}$, $h^{\mathrm{rt}}$, and
$h^{\mathrm{late}}$ denote deterministic arrivals, ray-traced reflections,
and the parametric late tail, respectively. The transition weights satisfy
$w_{\mathrm{rt}}^2[n]+w_{\mathrm{late}}^2[n]=1$ within the overlap region,
providing an equal-power transition between the traced and synthesized
responses. The bandwise responses are finally reconstructed into a full-band
multichannel RIR. This decomposition preserves important geometric arrivals
without requiring dense high-order path tracing throughout the complete
reverberation decay~\cite{savioja2015overview,vorlander2008auralization}.

\subsection{Flexible Source and Microphone Configurations}
\label{sec:flexible_configurations}

Source and receiver configurations are specified independently of the static
acoustic scene. A source state contains its position, orientation,
directivity, emitted signal, and optional motion trajectory. AcousticAgent
supports omnidirectional and directional sources, including common first-order
patterns and focused radiation patterns. Moving sources are represented by
acoustic keyframes sampled along a user-defined trajectory.

Receiver configurations specify microphone positions, orientations, and
directional or binaural responses. The same interface supports a single
microphone, a binaural receiver, a compact microphone array, or distributed
microphones placed independently across different rooms. Array elements share
a common local coordinate system, whereas distributed microphones can have
arbitrary positions and orientations, enabling both conventional array
processing and household-scale multi-device sensing.

For a trajectory containing $K$ source states and a receiver configuration
with $Q$ channels, AcousticAgent generates
\begin{equation}
    \mathcal{H}
    =
    \left\{h_{q,k}[n]\right\}
    \in
    \mathbb{R}^{K\times Q\times L},
    \label{eq:rir_tensor}
\end{equation}
where $L$ denotes the RIR length. Static-source simulation is the special case
$K=1$. Because the scene geometry, materials, and room connectivity remain
unchanged, only source--receiver-dependent propagation needs to be recomputed
when varying source trajectories or microphone layouts. The generated RIRs
can subsequently be convolved with task-specific dry signals to synthesize
static or time-varying multichannel reverberant audio.

\begin{figure}[t]
    \centering
    \includegraphics[width=\linewidth]{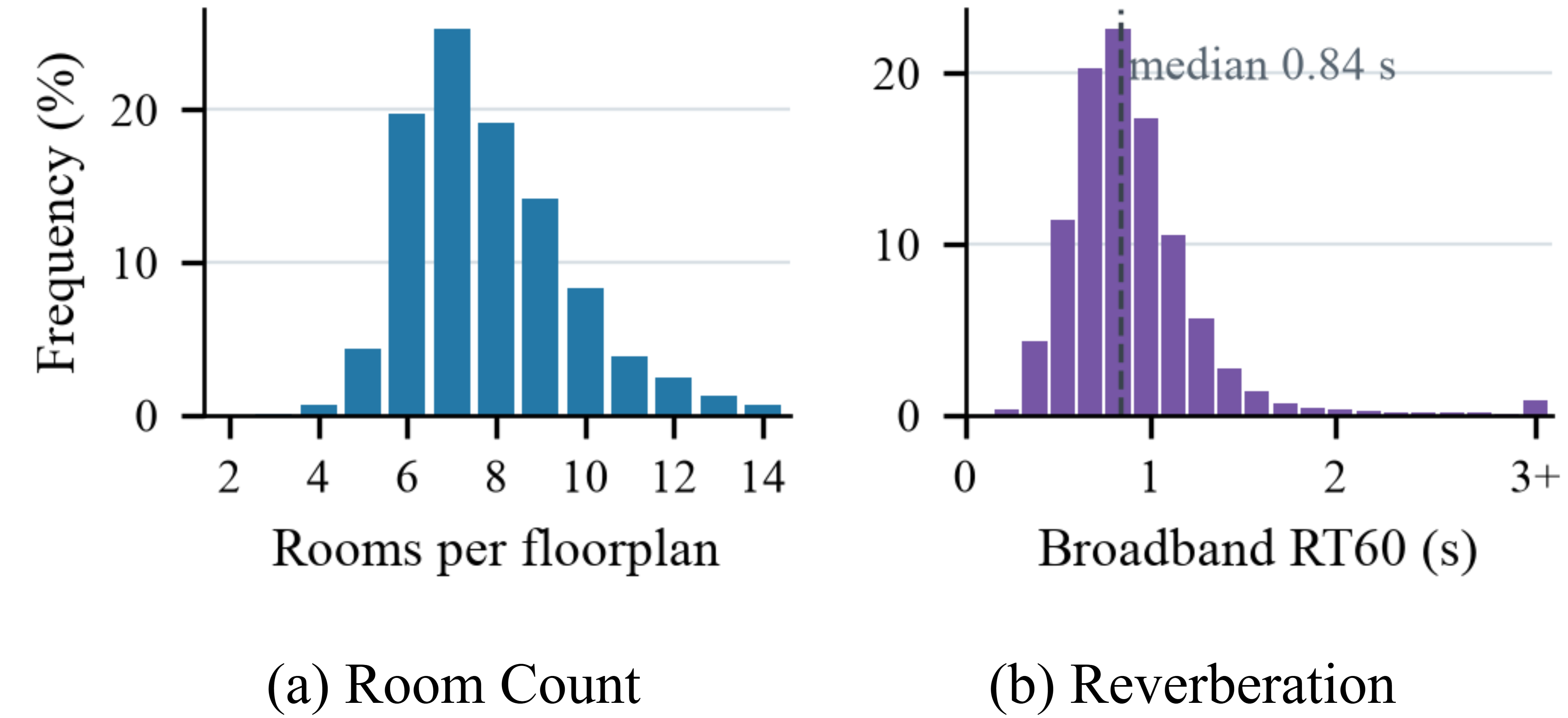}
    \caption{Structural and acoustic statistics of FP-RIR.
    (a) Distribution of the number of rooms per residential floorplan.
    (b) Distribution of broadband $RT_{60}$ values estimated from the generated RIRs; values exceeding $3.0\,\mathrm{s}$ are grouped into the final bin.}
    \label{fig:fprir_distributions}
\end{figure}

\subsection{FP-RIR Dataset}
\label{sec:fprir}

Using AcousticAgent, we construct \textbf{FP-RIR}, a floorplan-based multichannel RIR dataset for far-field and multi-room speech processing. FP-RIR contains 15,376 validated residential floorplans from ResPlan~\cite{abouagour2025resplan}, comprising 120,150 rooms and 107,595 open interior portals. For each floorplan, six-band acoustic materials are assigned according to scene semantics, and source and receiver positions are sampled from collision-free regions. For each floorplan, we generate two within-room monophonic configurations and up to three cross-room configurations recorded by four-channel circular microphone arrays. Cross-room source--receiver pairs cover different topological distances and are retained only when they are connected through valid sequences of open portals.

As shown in Fig.~\ref{fig:fprir_distributions}, FP-RIR covers residential floorplans containing 2--14 rooms, with a median of 7 rooms. The broadband $RT_{60}$ estimated from the generated RIRs has a median of 0.84 s and a 95th percentile of 1.48 s, reflecting acoustic diversity arising from variations in room geometry, material assignment, and cross-room connectivity.

\section{Experiments}
\label{sec:experiments}

We evaluate AcousticAgent in terms of acoustic fidelity and simulation
efficiency, distributed whole-home source localization, and distributed speech
enhancement. We further examine the downstream effects of the distributed
front ends using frozen ASR and speaker verification models.

\subsection{Acoustic Simulation Validation}
\label{sec:simulation_validation}

\textbf{Agreement with measured acoustics.} We validate AcousticAgent against dEchorate measurements
~\cite{dicarlo2021dechorate} using ten source--microphone configurations
selected through distance-stratified sampling. The documented room geometry
is reconstructed, and a fixed material configuration is applied across all
positions. We compare the direct-to-reverberant ratio (DRR) and reverberation
time ($RT_{60}$), averaged over 500\,Hz--4\,kHz, to evaluate spatial energy
balance and reverberation decay.

\begin{figure}[t]
    \centering
    \includegraphics[width=\linewidth]{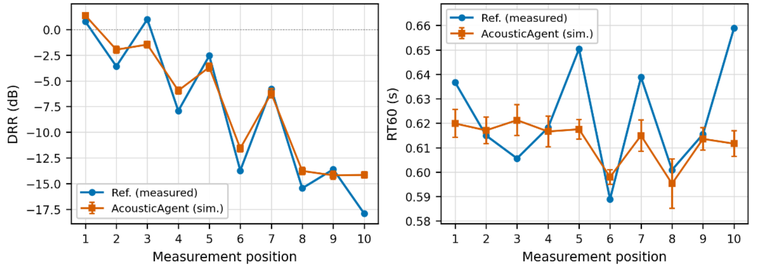}
    \caption{Validation against dEchorate measurements.
    Position-wise comparison of simulated and measured (a) DRR and
    (b) $RT_{60}$. Simulated results are averaged over three independent runs,
    with error bars indicating one standard deviation.}
    \label{fig:dechorate_drr_rt60}
\end{figure}

As shown in Fig.~\ref{fig:dechorate_drr_rt60}, AcousticAgent closely follows
the measured position-dependent DRR variation, achieving an MAE of
1.638\,dB and a Pearson correlation of $r=0.971$. For $RT_{60}$, the simulator
obtains an MAE of 0.016\,s and a mean relative error of 2.46\%. The close
agreement in both metrics indicates that AcousticAgent captures the
direct-to-reverberant energy balance and overall reverberation decay under
matched scene conditions.

\textbf{Simulation efficiency.} We evaluate the end-to-end RIR-generation throughput of the Numba and CUDA
backends on four furnished and unfurnished floorplan scenes with five or ten
rooms. The three settings range from 8,192 rays and 32 bounces to 131,072 rays
and 96 bounces, and runtime includes ray tracing, response construction, and
post-processing. Experiments are conducted using an Intel Xeon Gold 6346 CPU
and a single NVIDIA RTX 4090 GPU. Across the evaluated settings, Numba achieves
5.80--12.86 RIRs per second, while CUDA achieves 12.29--14.47 RIRs per second,
supporting efficient large-scale RIR generation.

% \begin{table}[t]
%     \centering
%     \small
%     \setlength{\tabcolsep}{4.5pt}
%     \renewcommand{\arraystretch}{1.08}
%     \caption{\textbf{End-to-end RIR-generation throughput under different
%     simulation settings.} Throughput is measured in generated RIRs per second.}
%     \label{tab:simulation_speed}
%     \begin{tabular}{@{}lrrrr@{}}
%         \toprule
%         \textbf{Setting} &
%         \textbf{Rays} &
%         \textbf{Bounces} &
%         \textbf{Numba} &
%         \textbf{CUDA} \\
%         \midrule
%         Preview
%             & 8,192   & 32 & 12.86 & 14.47 \\
%         Simulation
%             & 32,768  & 64 &  9.09 & 14.13 \\
%         Reference
%             & 131,072 & 96 &  5.80 & 12.29 \\
%         \bottomrule
%     \end{tabular}
% \end{table}

\begin{figure*}[t]
    \centering
    \includegraphics[width=\textwidth]{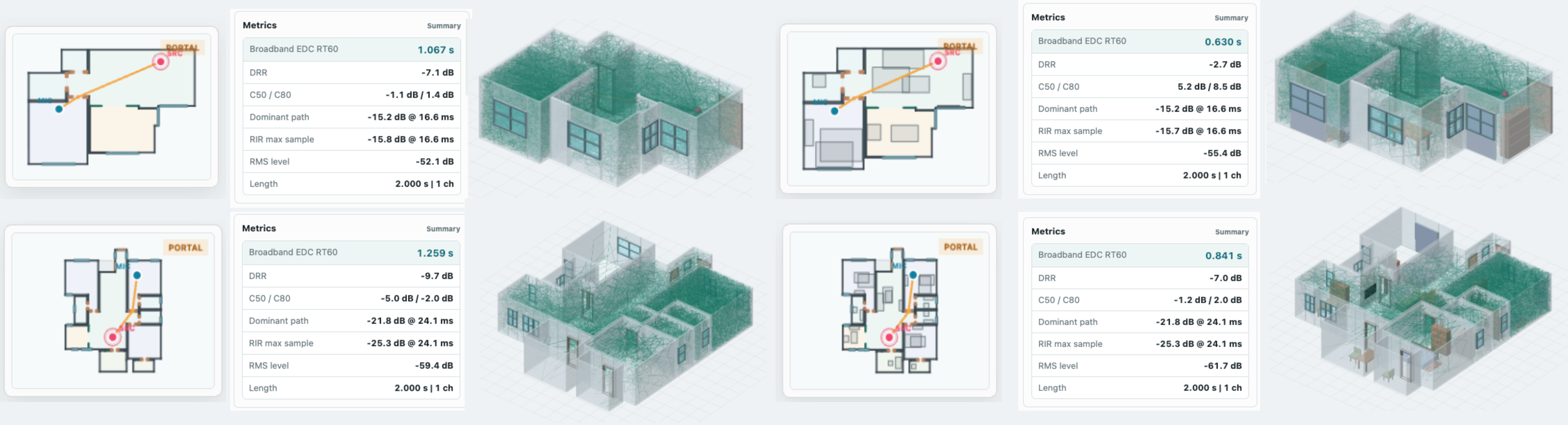}
    \caption{Qualitative examples of floorplan-aware acoustic propagation. For each residential layout, the 2-D view shows the source and receiver locations together with a valid cross-room portal sequence. The corresponding 3-D views illustrate sampled propagation rays in the unfurnished and furnished scenes, highlighting the effects of room connectivity, geometry, and furniture on acoustic propagation.}
    \label{fig:floorplan_acoustic_examples}
\end{figure*}

\textbf{Qualitative scene analysis.} Figure~\ref{fig:floorplan_acoustic_examples} presents furnished and
unfurnished variants of two residential layouts. Within each layout, the
source and receiver configurations are kept fixed to isolate the acoustic
effects of furniture and scene geometry. The examples illustrate how AcousticAgent jointly models room connectivity,
three-dimensional geometry, and material-dependent surface interactions.
Cross-room propagation follows valid portal routes, while furniture introduces
additional occlusion, reflection, scattering, and absorption. These changes
alter both the spatial distribution of propagation paths and the resulting
broadband reverberation characteristics.

\subsection{Distributed Whole-Home Acoustic Processing}
\label{sec:distributed_processing}

We investigate whether spatially distributed microphones provide more
consistent whole-home coverage than a fixed single-room array. We consider
two tasks: source localization and speech enhancement, both evaluated on
held-out residential floorplans with different numbers of rooms.

\subsubsection{Floorplan-Aware Source Localization}
\label{sec:distributed_localization}

\textbf{Experimental setup.} We evaluate 90 floorplans, with ten layouts for each room count from 4 to 12.
Three source positions are sampled in every room, resulting in 2,160 cases per
acoustic condition. A disjoint set of 45 floorplans is used only for
hyperparameter calibration.

For each source position, AcousticAgent generates an RIR to every microphone
channel in the whole-home deployment, producing a globally synchronized
multichannel observation across all rooms. A configuration with one microphone
per room therefore contains $R$ channels, whereas deploying a four-channel
array in each room contains $4R$ channels, where $R$ is the number of rooms.

Far-field speech is mixed with a spatially positioned MUSAN interference source
at SNRs of 20, 10, and 5\,dB. The target and interference signals are propagated
to all microphone channels using their corresponding RIRs. Independent device
noise with an equivalent sound pressure level of 30\,dB is then added to each
channel. We report source-room classification accuracy.

\textbf{Baselines.} The \emph{living-room array} applies SRP-PHAT using a single four-channel
circular array placed in the living room. \emph{GCC-Euclidean} and
\emph{GCC-FP} deploy one microphone per room and match GCC-PHAT delay
estimates using Euclidean and floorplan-aware propagation distances,
respectively. \emph{SRP-Euclidean} and \emph{SRP-FP} use the corresponding
steering models within the broadband SRP-PHAT objective.

The higher-channel \emph{distributed arrays} baseline places a four-channel
array in every room and combines inter-array TDOA estimates with DOA estimates
from the three arrays receiving the strongest signals.

\textbf{Floorplan-aware steering.} Euclidean steering assumes unobstructed free-field propagation and can be
inaccurate when walls block the direct path. We therefore construct a
visibility graph from the room polygons and open portals. The floorplan-aware
distance between candidate position $\mathbf{x}$ and microphone
$\mathbf{m}_i$ is defined as

\begin{equation}
    d_i^{\mathrm{FP}}(\mathbf{x})
    =
    \min_{\pi\in\mathcal{P}(\mathbf{x},\mathbf{m}_i)}
    L(\pi),
    \label{eq:floorplan_path_distance}
\end{equation}

where $\mathcal{P}(\mathbf{x},\mathbf{m}_i)$ denotes the set of geometrically
valid portal paths and $L(\pi)$ is the length of path $\pi$. The corresponding
TDOA is

\begin{equation}
    \tau_{ij}^{\mathrm{FP}}(\mathbf{x})
    =
    \frac{
        d_i^{\mathrm{FP}}(\mathbf{x})
        -
        d_j^{\mathrm{FP}}(\mathbf{x})
    }{c},
    \label{eq:floorplan_tdoa}
\end{equation}

where $c$ is the speed of sound. GCC-FP matches the estimated delays to
Eq.~\eqref{eq:floorplan_tdoa}, while SRP-FP uses the same delays in the
broadband SRP-PHAT objective.

\begin{table}[t]
    \centering
    % \small
    \setlength{\tabcolsep}{1mm}
    \renewcommand{\arraystretch}{1.08}
    \begin{tabular}{@{}llccc@{}}
        \toprule
        \textbf{Method} & \textbf{Mics} &
        \textbf{20 dB} & \textbf{10 dB} & \textbf{5 dB} \\
        \midrule
        Living-room array
            & $1{\times}4$
            & 18.9
            & 20.1
            & 18.1 \\
        GCC-Euclidean
            & $R{\times}1$
            & 81.9
            & 80.6
            & 76.5 \\
        GCC-FP
            & $R{\times}1$
            & 92.5
            & 89.9
            & 84.0 \\
        SRP-Euclidean
            & $R{\times}1$
            & 77.7
            & 75.0
            & 70.6 \\
        SRP-FP
            & $R{\times}1$
            & 81.8
            & 80.0
            & 76.8 \\
        Distributed arrays
            & $R{\times}4$
            & \textbf{94.4}
            & \textbf{90.0}
            & \textbf{85.2} \\
        \bottomrule
    \end{tabular}
    \caption{Whole-home source-room localization under spatial music
    interference. $R$ denotes the number of rooms. Results are reported as
    room classification accuracy (\%).}
    \label{tab:distributed_localization_musan}
\end{table}

\textbf{Results.} As shown in Table~\ref{tab:distributed_localization_musan}, the fixed
living-room array performs poorly for whole-home source-room localization.
With one microphone per room, GCC-FP achieves 92.5\%, 89.9\%, and 84.0\%
accuracy at 20, 10, and 5\,dB, respectively, approaching the distributed-array
baseline while using only one quarter of the channels. Floorplan-aware
propagation consistently outperforms Euclidean steering, demonstrating the
benefit of incorporating room connectivity and valid portal paths.

\subsubsection{Whole-Home Speech Enhancement}
\label{sec:source_level_whole_home}

\textbf{Experimental setup.} We follow the floorplan, source, interference, and microphone-deployment
settings used in the localization experiment. 
% # On the same 90 test floorplans,
% one target position is sampled in each room, yielding 720 cases. 
The main difference is the signal-mixing protocol: before RIR convolution, target
speech, spatial MUSAN music, and independent device noise are calibrated to
nominal source levels of 65, 55, and 30\,dB SPL, respectively. The
receiver-side SNR is not normalized and therefore varies naturally with room
geometry, source distance, and cross-room connectivity.

We compare a fixed $1{\times}4$ living-room array, four channels selected from
the $R{\times}1$ deployment using GCC-Euclidean or GCC-FP, and twelve channels
from the three selected arrays in the $R{\times}4$ deployment. All
configurations use the same reference-channel multichannel Wiener filter
(MWF). The target and noise covariance matrices are estimated from
simulator-separated signal components; the results therefore represent a
target-aware upper bound for comparing microphone deployments rather than a
blind enhancement system. Wet PESQ and STOI are computed against the
reverberant target at the reference channel and macro-averaged over
floorplans.

\begin{figure}[t]
    \centering
    \includegraphics[width=\columnwidth]{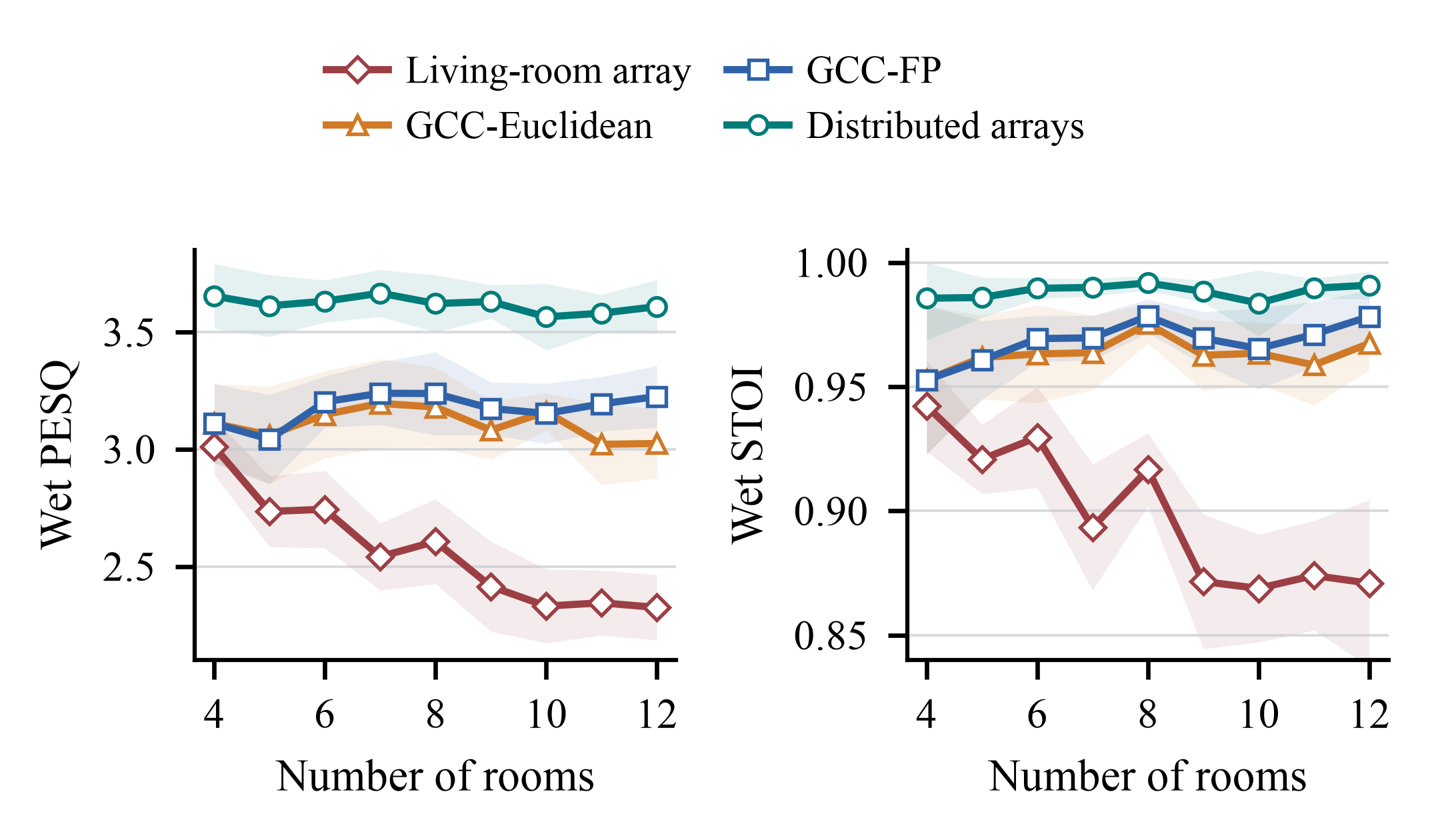}
    \caption{Whole-home speech enhancement across residential floorplan sizes.
    Wet PESQ (left) and Wet STOI (right) are reported as functions of the number
    of rooms. Lines denote floorplan-level macro averages, and shaded regions
    indicate 95\% confidence intervals over ten floorplans per group.}
    \label{fig:source_level_wet_metrics}
\end{figure}

\textbf{Results.} As shown in Table~\ref{tab:source_level_mwf}, distributed arrays achieve the
best overall performance, with a PESQ of 3.619 and a STOI of 0.988. Selecting
distributed microphones with GCC-FP also consistently outperforms the fixed
living-room array and provides a modest improvement over GCC-Euclidean.
Figure~\ref{fig:source_level_wet_metrics} further shows that the fixed array
degrades as floorplan size increases, whereas distributed arrays maintain
stable enhancement performance across different numbers of rooms.

\begin{table}[t]
    \centering
    \setlength{\tabcolsep}{1mm}
    \renewcommand{\arraystretch}{1.08}
    \begin{tabular}{@{}lccc@{}}
        \toprule
        \textbf{Method} & \textbf{Mics} &
        \textbf{PESQ} & \textbf{STOI} \\
        \midrule
        Living-room array
            & $1{\times}4$
            & 2.561 & 0.899 \\
        GCC-Euclidean
            & $R{\times}1{\rightarrow}4$
            & 3.109 & 0.963 \\
        GCC-FP
            & $R{\times}1{\rightarrow}4$
            & 3.175 & 0.968 \\
        Distributed arrays
            & $R{\times}4{\rightarrow}12$
            & \textbf{3.619} & \textbf{0.988} \\
        \bottomrule
    \end{tabular}
    \caption{Whole-home speech enhancement with calibrated source levels.
    $R$ denotes the number of rooms, and $\rightarrow K$ indicates that
    $K$ channels are selected for multichannel Wiener filtering (MWF).}
    \label{tab:source_level_mwf}
\end{table}

\begin{table}[t]
    \centering
    {
    \setlength{\tabcolsep}{1mm}
    \renewcommand{\arraystretch}{1.10}
    \begin{tabular}{@{}lccc@{}}
        \toprule
        \textbf{Model} &
        \shortstack{\textbf{Living}\\\textbf{array}} &
        \shortstack{\textbf{GCC}\\\textbf{FP}} &
        \shortstack{\textbf{Distributed}\\\textbf{arrays}} \\
        \midrule
        Whisper Tiny
            & 70.33
            & 22.93
            & \textbf{19.91} \\
        Whisper Small
            & 34.84
            & 9.27
            & \textbf{7.63} \\
        \bottomrule
    \end{tabular}
    }
    \caption{ASR performance after whole-home speech enhancement.
    Word error rates (WER, \%) are reported for frozen Whisper models with
    different MWF front ends. Lower is better.}
    \label{tab:distributed_downstream_asr}
\end{table}

\begin{table}[t]
    \centering
    {
    \setlength{\tabcolsep}{1mm}
    \renewcommand{\arraystretch}{1.10}
    \begin{tabular}{@{}lccc@{}}
        \toprule
        \textbf{Model} &
        \shortstack{\textbf{Living}\\\textbf{array}} &
        \shortstack{\textbf{GCC}\\\textbf{FP}} &
        \shortstack{\textbf{Distributed}\\\textbf{arrays}} \\
        \midrule
        CAM++
            & 4.8 / 0.330
            & 2.2 / 0.086
            & \textbf{1.8} / \textbf{0.064} \\
        ERes2Net
            & 6.2 / 0.278
            & 1.8 / 0.158
            & \textbf{1.6} / \textbf{0.146} \\
        \bottomrule
    \end{tabular}
    }
    \caption{Speaker-verification performance after whole-home speech
    enhancement using frozen models and clean enrollment utterances.
    Results are reported as EER (\%) / minDCF, with
    $P_{\mathrm{tar}}=0.01$ and
    $C_{\mathrm{miss}}=C_{\mathrm{fa}}=1$. Lower is better.}
    \label{tab:distributed_downstream_sv}
\end{table}

\subsection{Downstream Benefits of Distributed Front Ends}
\label{sec:distributed_downstream}

\textbf{Experimental setup.} We reuse the test scenes, source-level-calibrated mixtures, microphone
deployments, and MWF front ends from the whole-home speech enhancement
experiment. The enhanced signals are evaluated using frozen Whisper Tiny and
Whisper Small models for ASR~\cite{radford2023whisper}, and frozen CAM++~\cite{wang2023campp} and ERes2Net~\cite{chen2023eres2net} models for speaker
verification. We use 1,000 LibriSpeech~\cite{panayotov2015librispeech} test-clean utterances for ASR and
construct 1,000 VoxCeleb1~\cite{nagrani2017voxceleb} verification trials with clean enrollment.

\textbf{Results.}

Distributed arrays achieve the lowest WER for both ASR models
(Table~\ref{tab:distributed_downstream_asr}), reducing WER by 71.7--78.1\%
relative to the fixed living-room array. GCC-FP also provides substantial
improvements under the smaller four-channel budget, showing that
floorplan-aware microphone selection benefits downstream recognition.

For speaker verification, both distributed configurations substantially
reduce EER and minDCF compared with the fixed array
(Table~\ref{tab:distributed_downstream_sv}). Distributed arrays obtain the
best overall results, indicating that spatially distributed capture improves
both speech recognition and speaker verification without downstream-model
adaptation.

\section{Conclusion}
\label{sec:conclusion}

In this paper, we present AcousticAgent, a floorplan-centric platform for constructing editable multi-room acoustic scenes, together with FP-RIR, a large-scale multichannel RIR dataset. AcousticAgent combines semantic material parameterization, portal-aware propagation, Monte Carlo ray tracing, and parametric late reverberation under flexible source and microphone configurations. Validation against measured RIRs demonstrates reliable estimation of DRR and reverberation characteristics. Distributed benchmarks show that one microphone per room can approach the source-room localization performance of array-per-room systems with fewer channels, while distributed arrays provide the most consistent performance in speech enhancement and downstream ASR and speaker verification. Together, AcousticAgent and FP-RIR provide a scalable and reproducible testbed for residential distributed speech processing. Future work will investigate measured multi-room calibration, asynchronous devices, dynamic portal states, and moving speakers.

\bibliographystyle{IEEEbib}
\bibliography{refs}

\newpage

\section*{Appendix}

\setcounter{section}{0}

\section{Floorplan-Aware Acoustic Simulation}
\label{sec:supp_acoustic_solver}

This section details the acoustic formulation of AcousticAgent.  Given a
geometric scene, frequency-dependent material parameters, and source/receiver
states, the solver constructs the multichannel room impulse response (RIR) as
\begin{align}
    h_q[n]
    & =
    \sum_{b\in\mathcal{B}}
    \left(
        h^{\mathrm{det}}_{q,b}[n]
        + h^{\mathrm{hyb}}_{q,b}[n]
    \right), \\
    \mathcal{B}
    & = \{125,250,500,1000,2000,4000\}\ \mathrm{Hz},
    \label{eq:supp_rir_decomposition}
\end{align}
where $q$ indexes the receiver channel and $b$ indexes one of six octave bands.
The deterministic term contains sparse direct, transmitted, portal-routed,
and edge-diffracted arrivals.  The hybrid term combines Monte Carlo early/mid
reflections with a dense parametric late tail generated by a 16-delay feedback
delay network (FDN).

\subsection{Scene and Material Parameterization}
\label{sec:supp_scene_parameterization}

The scene comprises piecewise-planar room boundaries, open portals, semantic
furniture proxies, a source at $\mathbf{s}$, and a receiver at $\mathbf{r}$.
Each surface instance is assigned band-dependent absorption
$\alpha_{m,b}$, scattering $\sigma_{m,b}$, and transmission loss
$\mathrm{TL}_{m,b}$ in decibels.  The corresponding pressure-domain
reflection and transmission magnitudes are
\begin{equation}
    \rho_{m,b}=\sqrt{\max(0,1-\alpha_{m,b})},
    \qquad
    \tau_{m,b}=10^{-\mathrm{TL}_{m,b}/20}.
    \label{eq:supp_material_coefficients}
\end{equation}
The source is represented by a weighted-dipole directivity model.  For source
forward direction $\mathbf{f}$ and unit propagation direction $\mathbf{u}$,
\begin{equation}
    D_s(\mathbf{u})
    =
    \left|
        (1-w)+w\,\mathbf{f}^{\mathsf T}\mathbf{u}
    \right|^p,
    \qquad 0\leq w\leq 1,
    \label{eq:supp_source_directivity}
\end{equation}
where $(w,p)$ parameterizes omnidirectional, cardioid, dipole, focused, and
intermediate patterns.  Ray--surface intersections may be accelerated by a
bounding volume hierarchy (BVH).  BVH traversal changes computational cost
only and leaves the acoustic solution unchanged.

\subsection{Deterministic Propagation}
\label{sec:supp_deterministic}

\textbf{Direct sound, occlusion, and transmission.}
Let $d_0=\lVert\mathbf{s}-\mathbf{r}\rVert_2$, $c$ be the speed of sound,
$d_{\min}$ the near-field distance clamp, and $\mu_b$ the air-absorption
coefficient in nepers per meter.  The direct delay and band gain are
\begin{align}
    t_0 &= d_0/c, \\
    g^{\mathrm{dir}}_b
    &=
    \frac{D_s(\mathbf{u}_0)}{\max(d_0,d_{\min})}
    \exp(-\mu_b d_0)
    \left[
        o+(1-o)\,T_b
    \right],
    \label{eq:supp_direct_gain}
\end{align}
where $o\in[0,1]$ is a visibility factor that permits binary geometric or soft
volumetric occlusion.  For an occluded segment, every distinct intersected
barrier contributes once,
\begin{equation}
    T_b=\prod_{m\in\mathcal{I}(\mathbf{s},\mathbf{r})}\tau_{m,b}.
    \label{eq:supp_barrier_transmission}
\end{equation}
An unobstructed path therefore has $o=1$.  A fully occluded path has $o=0$ and
contributes only through the accumulated material transmission term.

\textbf{Open-portal routing.}
For source and receiver positions in different rooms, open portals form a room
graph $\mathcal{G}_{\mathrm{room}}$.  A graph edge corresponding to portal $j$
has cost
\begin{equation}
    c_j = 1 + \frac{0.05}{\max(w_j,0.1)},
    \label{eq:supp_portal_graph_cost}
\end{equation}
where $w_j$ is its width in meters; this cost mildly favors wider openings.
Dijkstra search first determines a topologically valid portal sequence.
Within each traversed room, a visibility graph then connects the source,
receiver, portal-side points, and polygon vertices.  The resulting route
$\pi=(\mathbf{x}_0=\mathbf{s},\ldots,\mathbf{x}_{J+1}=\mathbf{r})$ is accepted
only if every segment is collision-free and crosses the prescribed open
apertures.  Its length and accumulated turning deviation are
\begin{align}
    d_{\pi} &= \sum_{j=0}^{J}
        \lVert\mathbf{x}_{j+1}-\mathbf{x}_{j}\rVert_2, \\
    \Theta_{\pi} &= \sum_{j=1}^{J}
        \cos^{-1}\!\left(
        \widehat{\mathbf{v}}_{j-1}^{\mathsf T}
        \widehat{\mathbf{v}}_{j}
        \right),
    \quad
    \mathbf{v}_{j}=\mathbf{x}_{j+1}-\mathbf{x}_{j}.
    \label{eq:supp_portal_geometry}
\end{align}
The portal-path gain is
\begin{equation}
    g^{\mathrm{portal}}_{\pi,b}
    =
    \frac{D_s(\widehat{\mathbf{v}}_0)}
         {\max(d_{\pi},d_{\min})}
    \exp(-\mu_b d_{\pi})
    \Psi_b(\Theta_{\pi})\,a_{\pi},
    \label{eq:supp_portal_gain}
\end{equation}
where $\Psi_b$ is the frequency-dependent path-deviation response defined
below.  Because portal geometry already constrains visibility, the default is
$a_{\pi}=1$, avoiding a second attenuation for the same opening.  When a
diffuse-aperture approximation is desired, the optional factor is
\begin{equation}
    a_{\pi}=\prod_{j=1}^{J}\sqrt{\eta_j},
    \qquad
    \eta_j=
    \frac{A_j}{A_j+2\pi d_j^-d_j^+},
    \label{eq:supp_portal_aperture}
\end{equation}
with portal area $A_j$ and incoming/outgoing distances $d_j^-$ and $d_j^+$.

\textbf{Edge diffraction.}
For a non-line-of-sight (NLOS) source--receiver pair within the same room, the
solver enumerates valid reflex boundary vertices and furniture edges.  It
retains at most $K_{\max}$ shortest candidates up to order $O_{\max}$.  For
edge sequence $\gamma$, distance $d_{\gamma}$ and accumulated angular
deviation $\Theta_{\gamma}$ are computed as in
Eq.~\eqref{eq:supp_portal_geometry}, giving
\begin{equation}
    g^{\mathrm{diff}}_{\gamma,b}
    =
    \frac{D_s(\widehat{\mathbf{v}}_0)}
         {\max(d_{\gamma},d_{\min})}
    \exp(-\mu_b d_{\gamma})
    \Psi_b(\Theta_{\gamma}).
    \label{eq:supp_diffraction_gain}
\end{equation}
We use a uniform-theory-of-diffraction (UTD)-inspired deviation model with
wedge index $\nu=2$, effective edge length $L_e=0.05$~m, and
$k_b=2\pi f_b/c$.  Define the shared phase term
\begin{equation}
    \phi(x)=
    \exp\!\left[
        \frac{\mathrm{i}\pi}{4}
        \sqrt{\frac{x}{x+1.4}}
    \right].
    \label{eq:supp_utd_phase}
\end{equation}
The transition function is then
\begin{equation}
    F(x)=
    \begin{cases}
    \sqrt{\pi x}\left(1-\dfrac{\sqrt{x}}
        {0.7\sqrt{x}+1.2}\right)
        \phi(x), & x<0.8,\\[6pt]
    \left(1-\dfrac{0.8}{(x+1.25)^2}\right)
        \phi(x), & x\geq 0.8.
    \end{cases}
    \label{eq:supp_utd_transition}
\end{equation}
Writing $\beta_\ell$ for the four UTD angular combinations,
$s_\ell\in\{+1,-1,+1,-1\}$, and $n_\ell$ for the nearest admissible wedge
integer, the raw coefficient is
\begin{align}
    D_b(\Theta)
    &=
    \frac{e^{-\mathrm{i}\pi/4}}
         {2\nu\sqrt{2\pi k_b}}
    \sum_{\ell=1}^{4}
    \cot\!\left(
        \frac{\pi+s_\ell\beta_\ell}{2\nu}
    \right)
    F\!\left(k_b L_e a_\ell\right), \\
    a_\ell
    &=2\cos^2\!\left(
        \pi\nu n_\ell-\frac{\beta_\ell}{2}
    \right),
    \qquad
    \alpha_i=0,\quad \alpha_d=\pi+\Theta.
    \label{eq:supp_utd_coefficient}
\end{align}
The response is normalized by its near-zero-deviation reference.  A $1/16$
cross-band floor prevents unrealistically deep band nulls:
\begin{align}
    R_b(\Theta)
    &=
    \frac{\min(1,|D_b(\Theta)|)}
         {\max(\min(1,|D_b(\varepsilon)|),\epsilon)}, \\
    \Psi_b(\Theta)
    &=
    \max\!\left(
        R_b(\Theta),\frac{\max_{b'}R_{b'}(\Theta)}{16}
    \right).
    \label{eq:supp_utd_normalization}
\end{align}
For a multi-edge candidate, $\Psi_b$ is evaluated once at the total deviation
$\Theta_{\gamma}$ rather than as a product of independent edge gains.  Portal
routing and diffraction have distinct roles: portal routing enforces legal
propagation through open apertures, whereas diffraction models bending around
explicit boundary or object edges.  For cross-room propagation, an additional
edge-diffraction path is not added for a connection already represented by a
portal route, thereby avoiding duplicated energy.

\textbf{Discrete path impulses.}
Let $\mathcal{P}_{\mathrm{det}}$ be the accepted direct, transmitted, portal,
and diffracted paths.  Their six-band impulse train is
\begin{equation}
    h^{\mathrm{det}}_b[n]
    =
    \sum_{p\in\mathcal{P}_{\mathrm{det}}}
    g_{p,b}\,kappa\!\left(n-f_s d_p/c\right),
    \label{eq:supp_discrete_impulses}
\end{equation}
where $\kappa$ is an energy-normalized, Hann-windowed sinc kernel that places
each arrival at a sub-sample delay.  Integer-delay rendering is retained as a
lower-cost option.

\subsection{Listener-Space Monte Carlo Reflections}
\label{sec:supp_early_reflections}

By acoustic reciprocity, the reflection solver launches $N_r$ rays uniformly
from the listener and performs next-event connection to the source.  At
surface hit $j$, a contribution is recorded only when the source--hit shadow
segment is visible.  Let $\ell_j$ denote the source-to-hit distance, $L_j$ the
accumulated listener-to-hit ray length, and $\alpha_{j,b}$ and $\sigma_{j,b}$
the local absorption and scattering coefficients.  The surviving path energy
before hit $j$ is
\begin{equation}
    Q_{j,b}=\prod_{u<j}(1-\alpha_{u,b})
    \label{eq:supp_path_survival}
\end{equation}
the energy surviving previous bounces.  The diffuse-plus-specular scattering
term is
\begin{equation}
    f_j
    =
    \frac{\sigma_j}{\pi}\cos\theta_{i,j}
    +
    \frac{m+2}{8\pi}(1-\sigma_j)
        \cos^m\theta_{h,j},
    \label{eq:supp_reflection_brdf}
\end{equation}
where $m$ controls the concentration of the specular lobe.  The resulting
next-event contribution in band $b$ is
\begin{equation}
    e_{j,b}
    =
    \frac{D_s(\mathbf{u}_{s,j})}{N_r\max(\ell_j,d_{\mathrm{irr}})^2}
    f_j\,(1-\alpha_{j,b})Q_{j,b}.
    \label{eq:supp_rt_energy_contribution}
\end{equation}
This expression combines the Monte Carlo ray weight $4\pi/N_r$ with source
irradiance $1/(4\pi\ell_j^2)$.  Relative to the direct arrival, its delay and
echogram bin are
\begin{align}
    \Delta t_j &= \frac{L_j+\ell_j-d_0}{c}, \\
    k_j &= \left\lfloor\frac{\Delta t_j}{\Delta t_{\mathrm{bin}}}\right\rfloor,
    \qquad
    E_b[k]=\sum_{j:k_j=k}e_{j,b}.
    \label{eq:supp_echogram_bins}
\end{align}
At each bounce, the outgoing direction is sampled from the diffuse lobe with
probability $\sigma_j$ and mirror-reflected otherwise.  This procedure yields
a six-band, time-binned echogram while retaining a bounded number of explicit
path segments for diagnostics and visualization.

The energy echogram is converted to a band-limited stochastic pressure
response.  For bin energy $E_b[k]$, the continuous pressure envelope is
\begin{equation}
    a_b(t)
    =
    \sqrt{\frac{\widetilde{E}_b(t)}{4\pi}}
    \exp(-\mu_b c t),
    \label{eq:supp_echogram_envelope}
\end{equation}
where $\widetilde{E}_b(t)$ linearly interpolates adjacent bins.  A seeded white
sequence $\xi[n]$ supplies a reproducible stochastic phase, and an eighth-order
crossover bank $\mathcal{F}_b\{\cdot\}$ confines each response to its octave
band:
\begin{equation}
    h^{\mathrm{tr}}_b[n]
    =
    \mathcal{F}_b\!\left\{
        a_b(n/f_s)\,\xi[n]
    \right\}.
    \label{eq:supp_traced_ir_reconstruction}
\end{equation}

\subsection{Decay Estimation and Coupled Rooms}
\label{sec:supp_decay_estimation}

Because $\mu_b$ is defined in the pressure domain, air absorption enters the
energy decay with exponent $2\mu_b$.  The Schroeder energy decay curve (EDC)
is
\begin{align}
    C_b[k]
    &=
    \sum_{u=k}^{K-1}
        E_b[u]\exp(-2\mu_b c t_u), \\
    L_b[k]
    &=
    10\log_{10}\!\left(
        \frac{C_b[k]}{C_b[0]}
    \right).
    \label{eq:supp_schroeder_edc}
\end{align}
A least-squares line $L_b(t)\approx\beta_b t+\zeta_b$ is fitted from
$-5$ to $-25$~dB and extrapolated to $-60$~dB:
\begin{equation}
    T_{60,b}=-\frac{60}{\beta_b},
    \qquad \beta_b<0.
    \label{eq:supp_rt60_fit}
\end{equation}
To identify coupled-room decay, a wider $-5$ to $-35$~dB interval is also
tested with a two-segment fit.  A double slope is accepted only when it improves
the Bayesian information criterion (BIC) by at least 10, the two equivalent
reverberation times differ by a factor of at least 1.35, and both segments
satisfy $R^2\geq0.95$.

For a connected floorplan, a single-room decay prior is insufficient.  We
therefore stabilize the late target using a coupled-room energy model.  Let
$V_i$ be room volume, $A_{i,b}$ its equivalent absorption area including
furniture, $A_i^{\mathrm{ext}}$ its exterior opening area, and $P_{ij}$ the
open-portal area between rooms $i$ and $j$.  The band-energy state
$\mathbf{e}_b(t)$ evolves as
\begin{equation}
    \frac{d\mathbf{e}_b(t)}{dt}
    =\mathbf{M}_b\mathbf{e}_b(t),
    \qquad
    \mathbf{e}_b(t)=\exp(\mathbf{M}_b t)\mathbf{e}_b(0),
    \label{eq:supp_coupled_energy_ode}
\end{equation}
with
\begin{align}
    [\mathbf{M}_b]_{ii}
    &=
    -\frac{c}{4V_i}
    \left(
        A_{i,b}+A_i^{\mathrm{ext}}+\sum_j P_{ij}
    \right), \\
    [\mathbf{M}_b]_{ij}
    &=
    \frac{cP_{ij}}{4V_j},
    \qquad i\neq j.
    \label{eq:supp_coupled_energy_matrix}
\end{align}
The initial state places unit energy in the source room.  Each negative
eigenvalue $\lambda_{b,m}$ of $\mathbf{M}_b$ defines a coupled decay mode with
equivalent reverberation time
\begin{equation}
    T_{60,b,m}
    =
    \frac{\ln(10^6)}{-\operatorname{Re}(\lambda_{b,m})}.
    \label{eq:supp_coupled_mode_rt60}
\end{equation}
The slowest mode contributing at least $1\%$ of the receiver-room energy sets
the cross-room late-tail prior, while the traced EDC provides a scene-specific
estimate.  Together they prevent the late tail from being determined solely
by either a local material prior or a noisy finite-ray fit.

\subsection{Sixteen-Delay Parametric Late Reverberation}
\label{sec:supp_fdn}

The finite-ray echogram becomes progressively sparse at late times.  We replace
this region with a dense, energy-matched $16$-line Hadamard FDN.  Let
$\mathbf{H}_{16}\mathbf{H}_{16}^{\mathsf T}=16\mathbf{I}$ and let $m_i$ denote
the delay, in samples, of line $i$.  The feedback state is
\begin{align}
    y_i[n] &= G_i(z)\,x_i[n-m_i], \\
    \mathbf{x}[n]
    &=
    \mathbf{b}\,\delta[n]
    +\frac{1}{4}\mathbf{H}_{16}\mathbf{y}[n],
    \label{eq:supp_fdn_state}
\end{align}
where $G_i(z)$ is a cascade of six compensated biquad sections.  Since
$\mathbf{H}_{16}/4$ is orthonormal, the unfiltered feedback mixing preserves
energy; frequency-dependent decay is introduced only by $G_i(z)$.  To suppress
coincident modes, delay lengths are deterministic prime powers, with
\begin{align}
    (p_1,\ldots,p_8)
    &=(2,3,5,7,11,13,17,19), \\
    (p_9,\ldots,p_{16})
    &=(23,29,31,37,41,43,47,53).
    \label{eq:supp_fdn_primes}
\end{align}
The delays are
\begin{align}
    m_i
    &=
    p_i^{\max\left(1,
        \operatorname{round}
        \left[
        \log_{p_i}(m_0+u_i)
        \right]
    \right)}, \\
    m_0
    &=\frac{0.15\times10\,f_s}{16},
    \qquad u_i\in\{0,\ldots,100\},
    \label{eq:supp_fdn_delays}
\end{align}
where seeded offset $u_i$ makes the network reproducible while decorrelating
its delay lines.  To realize target $T_{60,b}$, the round-trip pressure gain is
\begin{equation}
    g_{i,b}
    =
    \exp\!\left(
        -\frac{6.91\,m_i}{f_s T_{60,b}}
    \right).
    \label{eq:supp_fdn_feedback_gain}
\end{equation}
The compensated biquad cascade $G_i(z)$ is fitted to $g_{i,b}$ at the six band
centers.  The delay-line outputs are averaged and passed through four serial
all-pass diffusers with delays $\{225,341,441,556\}$ samples and coefficient
$0.5$, followed by multiband tone correction.  This stage increases temporal
density without changing the prescribed long-term decay.

Before rendering, the FDN is pre-rolled by $\max_i m_i$ plus the cumulative
all-pass delay, ensuring a populated state at the hybrid boundary.  Its level
is then calibrated to traced power at a per-band anchor $n_{a,b}$.  If
$P_{a,b}$ is the estimated traced power and
$\widetilde{h}^{\mathrm{fdn}}_b$ the unscaled FDN response, the target residual
energy and scale are
\begin{align}
    \mathcal{E}^{\star}_b
    &=
    \sum_{n=n_{a,b}}^{N-1}
    P_{a,b}
    \exp\!\left[
        -\frac{6\ln(10)(n-n_{a,b})}{f_sT_{60,b}}
    \right], \\
    s_b
    &=
    \sqrt{
        \frac{\mathcal{E}^{\star}_b}
        {\sum_{n=n_{a,b}}^{N-1}
        |\widetilde{h}^{\mathrm{fdn}}_b[n]|^2+\epsilon}
    }.
    \label{eq:supp_fdn_energy_calibration}
\end{align}

\textbf{Equal-power crossfade.}
For each band, let $n_{s,b}$ and $n_{e,b}$ denote the crossfade boundaries.
The nominal end is the configured hybrid transition; for coupled spaces it may
be replaced by the fitted EDC breakpoint.  Define
\begin{align}
    \lambda_b[n]
    &=
    \operatorname{clip}\!\left(
        \frac{n_{e,b}-n}{n_{e,b}-n_{s,b}},0,1
    \right), \\
    w^{\mathrm{tr}}_b[n]
    &=\sqrt{\lambda_b[n]}, \\
    w^{\mathrm{fdn}}_b[n]
    &=\sqrt{1-\lambda_b[n]}.
    \label{eq:supp_equal_power_weights}
\end{align}
Because $(w^{\mathrm{tr}}_b)^2+(w^{\mathrm{fdn}}_b)^2=1$, the crossfade
preserves expected power when the stochastic traced and FDN components are
approximately uncorrelated.  The hybrid reflection response is
\begin{equation}
    h^{\mathrm{hyb}}_b[n]
    =
    w^{\mathrm{tr}}_b[n]h^{\mathrm{tr}}_b[n]
    +
    w^{\mathrm{fdn}}_b[n]
        s_b\widetilde{h}^{\mathrm{fdn}}_b[n].
    \label{eq:supp_hybrid_reflection}
\end{equation}
Before $n_{s,b}$ the response is fully path traced; after $n_{e,b}$ it is fully
parametric.  The overlap provides a smooth transition without the energy dip
that can arise from a linear-amplitude crossfade.

\subsection{RIR Assembly and Receiver Rendering}
\label{sec:supp_receiver_rendering}

The reflection echogram uses time relative to the direct arrival and is shifted
by $n_0=\operatorname{round}(f_s d_0/c)$ during assembly.  The monaural RIR is
\begin{equation}
    h^{\mathrm{mono}}[n]
    =
    \sum_{b\in\mathcal{B}}
    \left(
        \mathcal{F}_b\{h^{\mathrm{det}}_b\}[n]
        +h^{\mathrm{hyb}}_b[n-n_0]
    \right).
    \label{eq:supp_mono_assembly}
\end{equation}
Bandwise and broadband reverberation times are re-estimated from the assembled
RIR using Eq.~\eqref{eq:supp_rt60_fit}.  Reported values therefore characterize
the rendered signal rather than being copied from a material or statistical
prior.

For a linear or circular microphone array with channel offsets
$\mathbf{o}_q$, the solver is evaluated at each physical channel position,
\begin{equation}
    h_q[n]
    =
    \mathcal{S}
    \left(
        \mathbf{s},\mathbf{r}+\mathbf{o}_q,
        \mathcal{G},\mathcal{M}
    \right)[n],
    \label{eq:supp_array_rendering}
\end{equation}
where $\mathcal{S}$ denotes the complete solver and $(\mathcal{G},\mathcal{M})$
the geometry and material parameters.  Solving at the physical channel
positions preserves inter-channel time and level differences; array channels
are not generated by duplicating a monaural response.

For binaural rendering, each deterministic path retains its arrival direction
$\boldsymbol{\Omega}_p$ and is filtered by a head-related impulse response
(HRIR) interpolated from SOFA data.  The reflection field additionally
accumulates first-order ambisonic (FOA) coefficients
\begin{equation}
    \mathbf{q}(\boldsymbol{\Omega})
    =
    [1,u_x,u_y,u_z]^{\mathsf T}.
    \label{eq:supp_foa_encoding}
\end{equation}
Let $d_{e,c}[n]$ be the SOFA-derived decoder filter for ear
$e\in\{L,R\}$ and FOA channel $c$.  The binaural RIR is
\begin{equation}
    h_e^{\mathrm{bin}}[n]
    =
    \sum_{p\in\mathcal{P}_{\mathrm{det}}}
        \left(h_p*r_e(\boldsymbol{\Omega}_p)\right)[n]
    +
    \sum_{c=0}^{3}
        \left(h_c^{\mathrm{FOA}}*d_{e,c}\right)[n].
    \label{eq:supp_binaural_rendering}
\end{equation}
The statistical FDN tail is injected into the omnidirectional FOA $W$ channel,
whereas traced early/mid reflections retain directional $X/Y/Z$ information.
HRIR interpolation combines the four nearest measurements in log-magnitude and
unwrapped-phase space.  When energy normalization is enabled, the common
binaural gain
\begin{equation}
    \gamma
    =
    \sqrt{
        \frac{\sum_n|h^{\mathrm{mono}}[n]|^2}
        {\frac{1}{2}\sum_{e\in\{L,R\}}\sum_n
        |h_e^{\mathrm{bin}}[n]|^2+\epsilon}
    }
    \label{eq:supp_binaural_normalization}
\end{equation}
is applied and capped at $24$~dB.

Finally, dry signal $x[n]$ is auralized by convolution,
\begin{equation}
    y_q[n]=(x*h_q)[n].
    \label{eq:supp_auralization}
\end{equation}
For multiple independent sources, each source is rendered using its own
position, directivity, and RIR before the receiver signals are summed.  Dynamic
scenes are represented by a sequence of RIRs evaluated at sampled source and
receiver keyframes, enabling time-varying convolution or framewise downstream
evaluation.

\subsection{VLM-Assisted Custom Floorplan Construction}
\label{sec:supp_vlm_floorplan}

Beyond the indexed residential corpus, AcousticAgent supports custom acoustic
scenes derived from user-provided floorplan images.  An external online
vision--language model (VLM) receives the image together with a
schema-constrained prompt and returns a structured description of room
polygons, walls, doors, windows, scale, and room connectivity.  The user may
calibrate the metric scale before submitting this description to
AcousticAgent.  The system then validates polygon topology, normalizes the
coordinate system, resolves shared boundaries, and converts valid openings
between adjacent rooms into acoustic portals.

The validated layout is lifted into a three-dimensional scene using the
specified room height.  Source and receiver positions, frequency-dependent
semantic materials, and optional furniture can subsequently be sampled or
edited through the same interface used for corpus floorplans.  Furniture may
be placed manually or generated by the semantic auto-placement module under
room-type, boundary-clearance, portal-clearance, and compactness constraints.
The resulting geometry is passed without modification to the floorplan-aware
solver described above.

Importantly, the VLM acts only as a geometric parser.  It neither predicts
acoustic coefficients nor synthesizes the RIR.  Material sampling, direct and
portal-mediated propagation, diffraction, Monte Carlo reflections, and the
parametric late tail remain deterministic components of AcousticAgent once
the structured scene and random seeds are fixed.  This separation makes the
custom-scene interface auditable while allowing a broad range of floorplan
styles to be simulated without manual geometric reconstruction.

\begin{figure}[t]
    \centering
    \includegraphics[width=\linewidth]{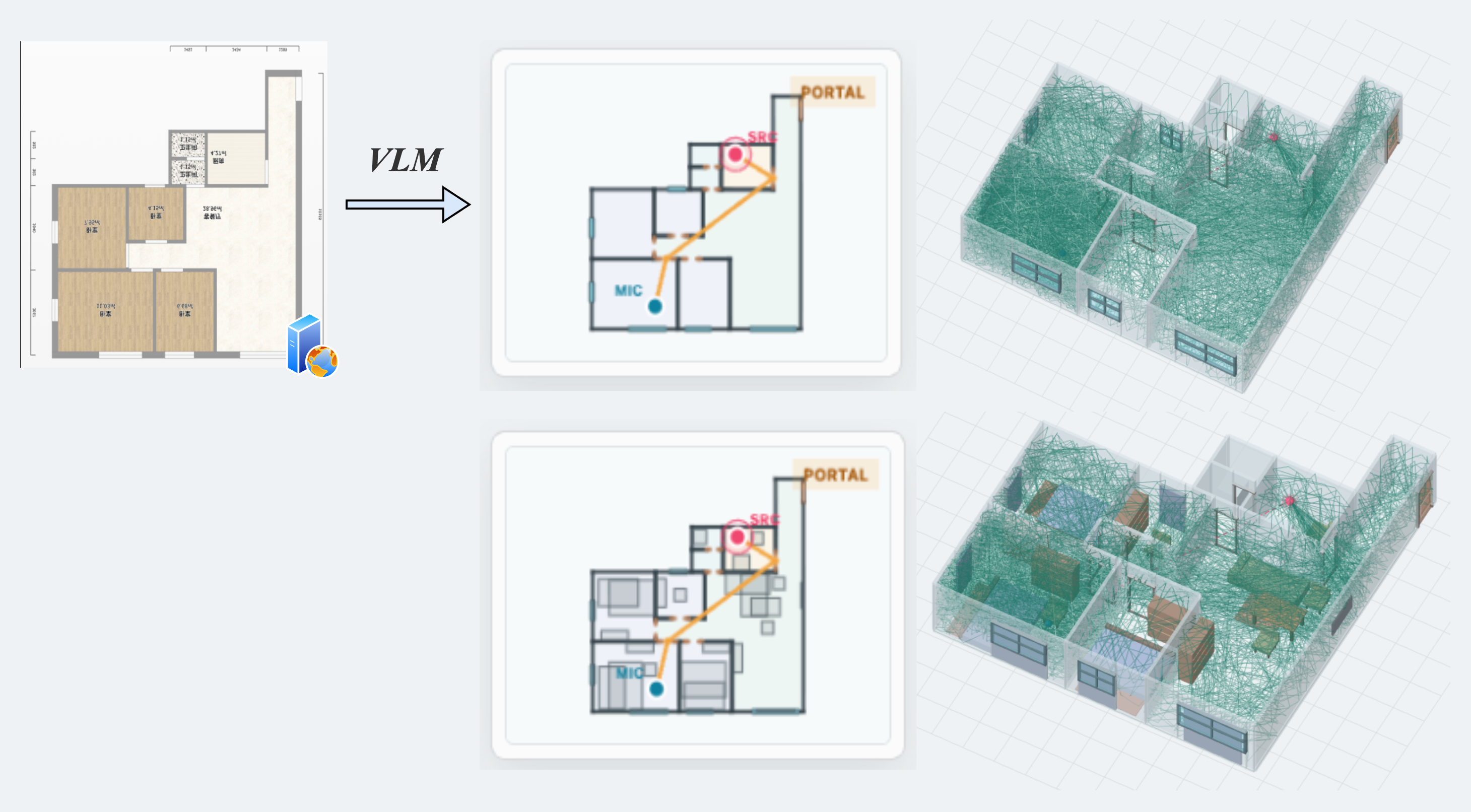}
    \caption{VLM-assisted construction of a custom acoustic scene.  A
    user-provided floorplan image is converted by an online VLM into a
    schema-constrained representation of room boundaries, openings, scale,
    and connectivity.  AcousticAgent validates and lifts this representation
    into simulation-ready geometry.  The upper row shows the geometry-only
    scene; the lower row shows the same layout after semantic furniture
    placement.  In the 2-D views, pink and blue markers denote the source and
    microphone, respectively, and the orange polyline denotes the valid
    portal route.  The corresponding 3-D views visualize the traced acoustic
    paths.}
    \label{fig:vlm_custom_floorplan}
\end{figure}

Figure~\ref{fig:vlm_custom_floorplan} illustrates the complete conversion from
a raster floorplan to unfurnished and furnished acoustic simulations.  The
same structured representation is reusable through the Python API, allowing
source/receiver configurations and material seeds to be varied without
invoking the VLM again.

\textbf{Prompt templates.}
The custom-scene interface is model agnostic: users may submit either a
floorplan image or a natural-language room specification to an external VLM
and paste the returned JSON into the workbench.  The exact prompt templates
are reproduced below.  Both prompts impose the same coordinate convention,
opening semantics, and schema so that the resulting scene passes through one
validation and compilation pipeline.

\begin{vlmprompt}{Prompt A: floorplan image to structured scene}
You are converting an attached residential floorplan image into AcousticAgent Floorplan JSON.

Preserve the image orientation exactly. Do not rotate, mirror, or vertically flip the plan. Trace the indoor outer boundary and every indoor room polygon. If a printed scale or dimensions are visible, use them. Otherwise set the outer width to 10 m, preserve the image aspect ratio, and record this assumption in provenance.scale_assumption. Do not invent an opening that is not visible.

Return exactly one JSON object and no Markdown. Follow these rules:
1. Use meters. Place the top-left of outer_boundary at [0, 0], with x increasing rightward and y downward.
2. Room polygons must not overlap, must remain inside outer_boundary, and together must cover the indoor region.
3. Supported room types are living, kitchen, bedroom, bathroom, storage, and balcony. Assign every room a unique id such as bedroom_0.
4. Represent each door, window, or wall-free opening as a two-point segment lying on the relevant room boundary.
5. An interior connection references exactly two room_ids and uses connection="interior_room". An entry door references one room and uses connection="outdoor_entry". An exterior window references one room and uses connection="outdoor_facade".
6. Set interior doors and wall-free openings to open=true. Set exterior entry doors and windows to open=false.
7. Use confidence values in [0, 1]. Use sill_height_m=0 for doors/openings and approximately 0.9 for windows unless the image specifies otherwise.
8. Preserve uncertain geometry conservatively and record assumptions in provenance.notes.

Required JSON shape:
{
  "schema_version": 1,
  "title": "...",
  "units": "m",
  "coordinate_system": "image_top_left",
  "height_m": 2.8,
  "wall_depth_m": 0.12,
  "outer_boundary": [[x, y], ...],
  "rooms": [
    {"id": "living_0", "type": "living", "corners": [[x, y], ...]}
  ],
  "openings": [
    {
      "id": "door_0",
      "type": "door",
      "room_ids": ["living_0", "bedroom_0"],
      "segment": [[x1, y1], [x2, y2]],
      "height_m": 2.1,
      "sill_height_m": 0.0,
      "connection": "interior_room",
      "open": true,
      "confidence": 0.9
    }
  ],
  "provenance": {
    "source": "vlm_image",
    "scale_assumption": null,
    "notes": []
  }
}

Before returning, verify polygon coverage, polygon overlap, boundary alignment, room references, and open-door connectivity.
\end{vlmprompt}

\begin{vlmprompt}{Prompt B: natural language to structured scene}
You are designing a residential floorplan for AcousticAgent from a user's description.

User request:
<USER_REQUEST>

Create a practical, compact, axis-aligned layout that satisfies the requested room counts and dimensions. If dimensions are missing, choose reasonable residential dimensions and record them in provenance.scale_assumption. Connect every indoor room to the living/circulation space through an open interior door or wall-free opening. Place plausible closed windows on exterior walls and include one outdoor entry door. Do not add rooms that contradict the request.

Return exactly one JSON object and no Markdown. Follow these rules:
 // same with Image prompt

Required JSON shape:
{
 // same with Image prompt
}

Before returning, verify polygon coverage, polygon overlap, boundary alignment, room references, and open-door connectivity.
\end{vlmprompt}

\begin{figure}[t]
    \centering
    \includegraphics[width=\linewidth]{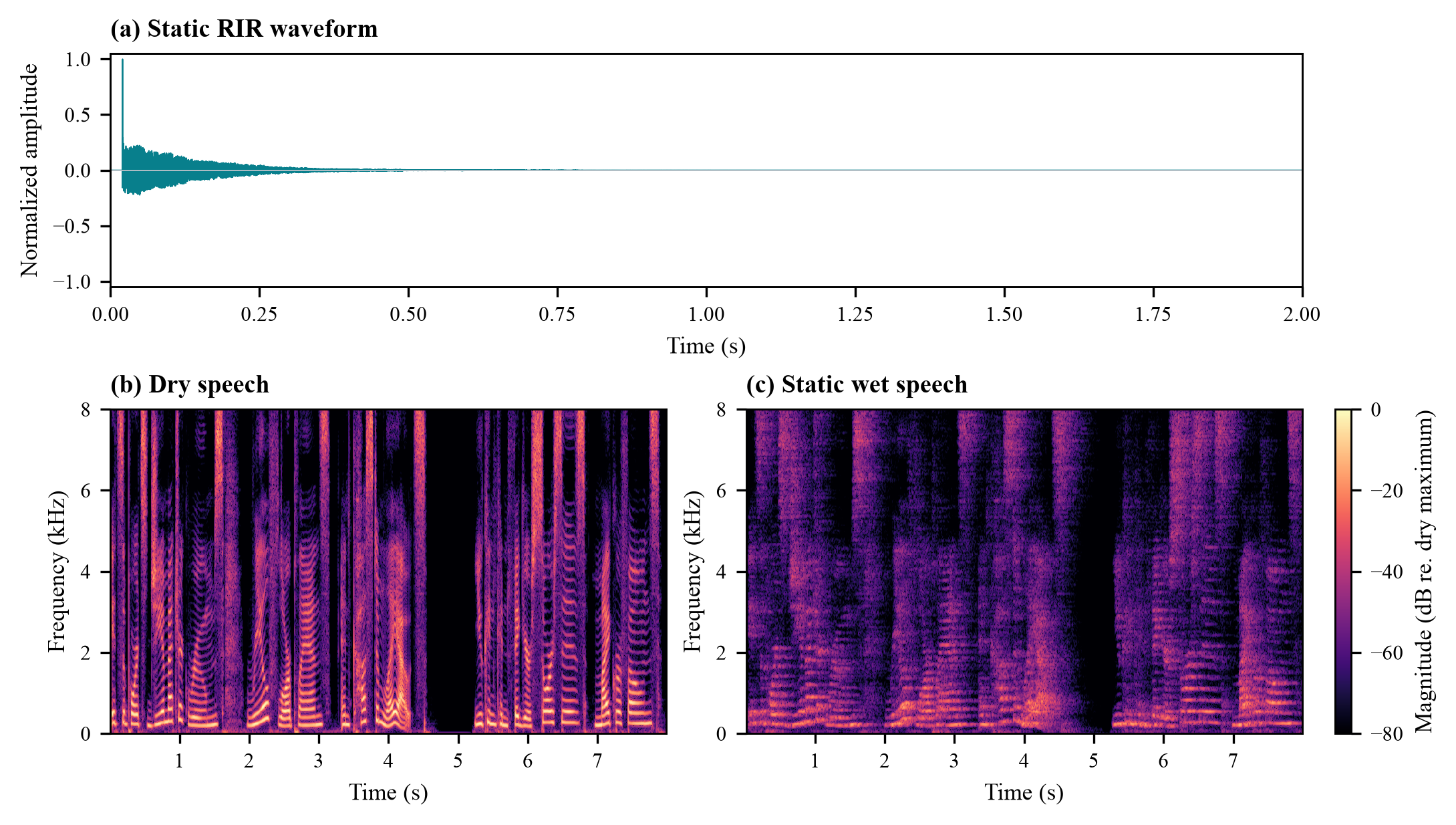}
    \caption{Static RIR and auralization example.  The source is located in
    the living room and the receiver in a bedroom of FloorPlan~\#5159, with a
    source--receiver distance of $6.79$~m and 24 semantic furniture objects.
    (a) Two-second monaural RIR generated at 16~kHz using the Simulation
    quality setting.  (b) Magnitude spectrogram of the dry speech signal.
    (c) Magnitude spectrogram after convolution with the static RIR.  The dry
    and wet spectrograms share the dry-speech maximum as their 0-dB reference;
    no independent level normalization is applied.}
    \label{fig:static_rir_auralization}
\end{figure}

\begin{figure}[t]
    \centering
    \includegraphics[width=\linewidth]{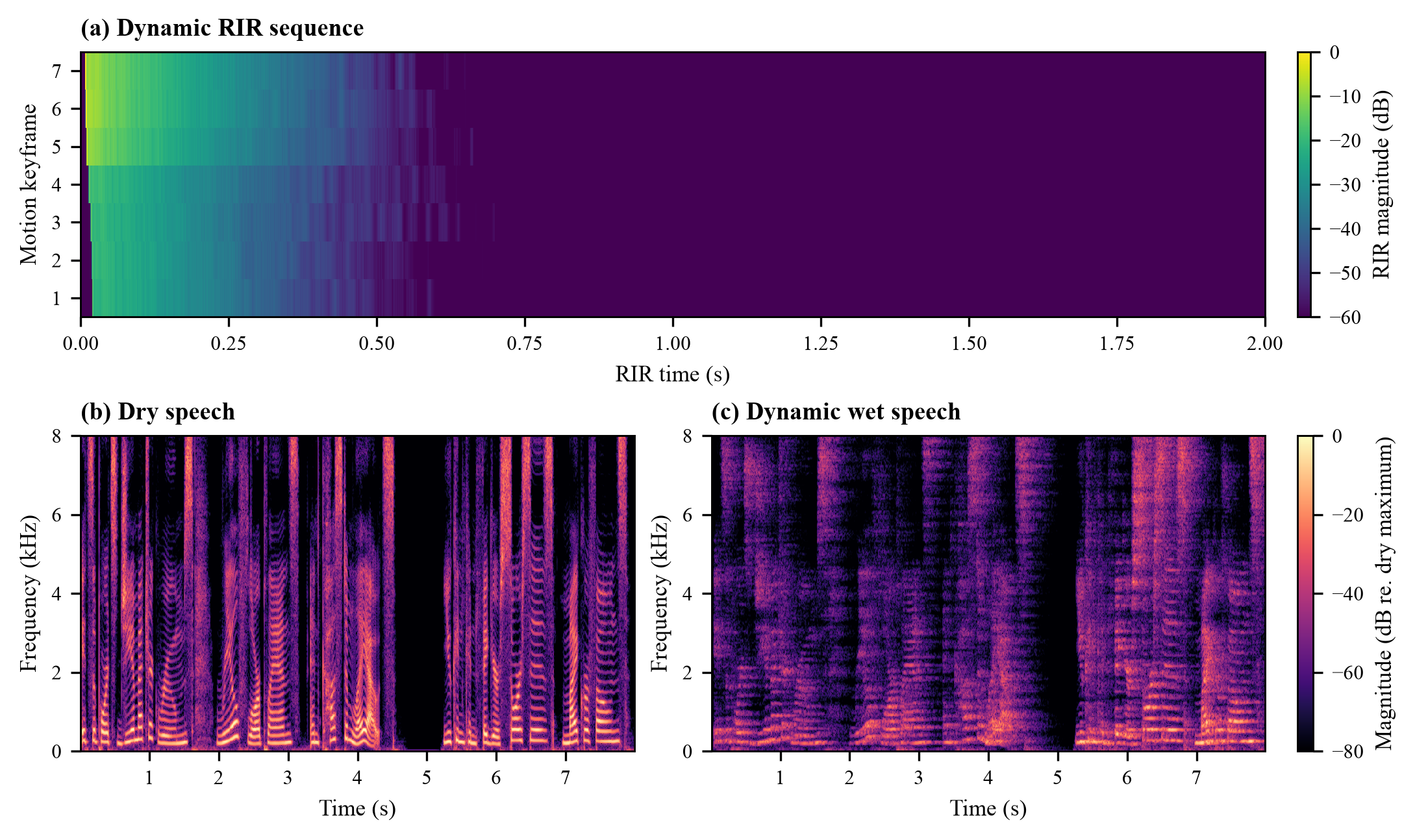}
    \caption{Dynamic RIR sequence and time-varying auralization.  The receiver
    follows a seven-keyframe, 4-m approach trajectory in FloorPlan~\#5159,
    reducing its source distance from $6.79$ to $2.94$~m.  (a) Smoothed RIR
    magnitude at every keyframe, using a common 0-dB reference across the
    sequence.  (b) Dry-speech magnitude spectrogram.  (c) Dynamic wet-speech
    spectrogram obtained by convolving the dry signal with each keyframe RIR
    and linearly interpolating adjacent rendered signals according to motion
    phase.  Panels (b) and (c) use a shared dry-speech amplitude reference.}
    \label{fig:dynamic_rir_auralization}
\end{figure}

\begin{figure}[t]
    \centering
    \includegraphics[width=\linewidth]{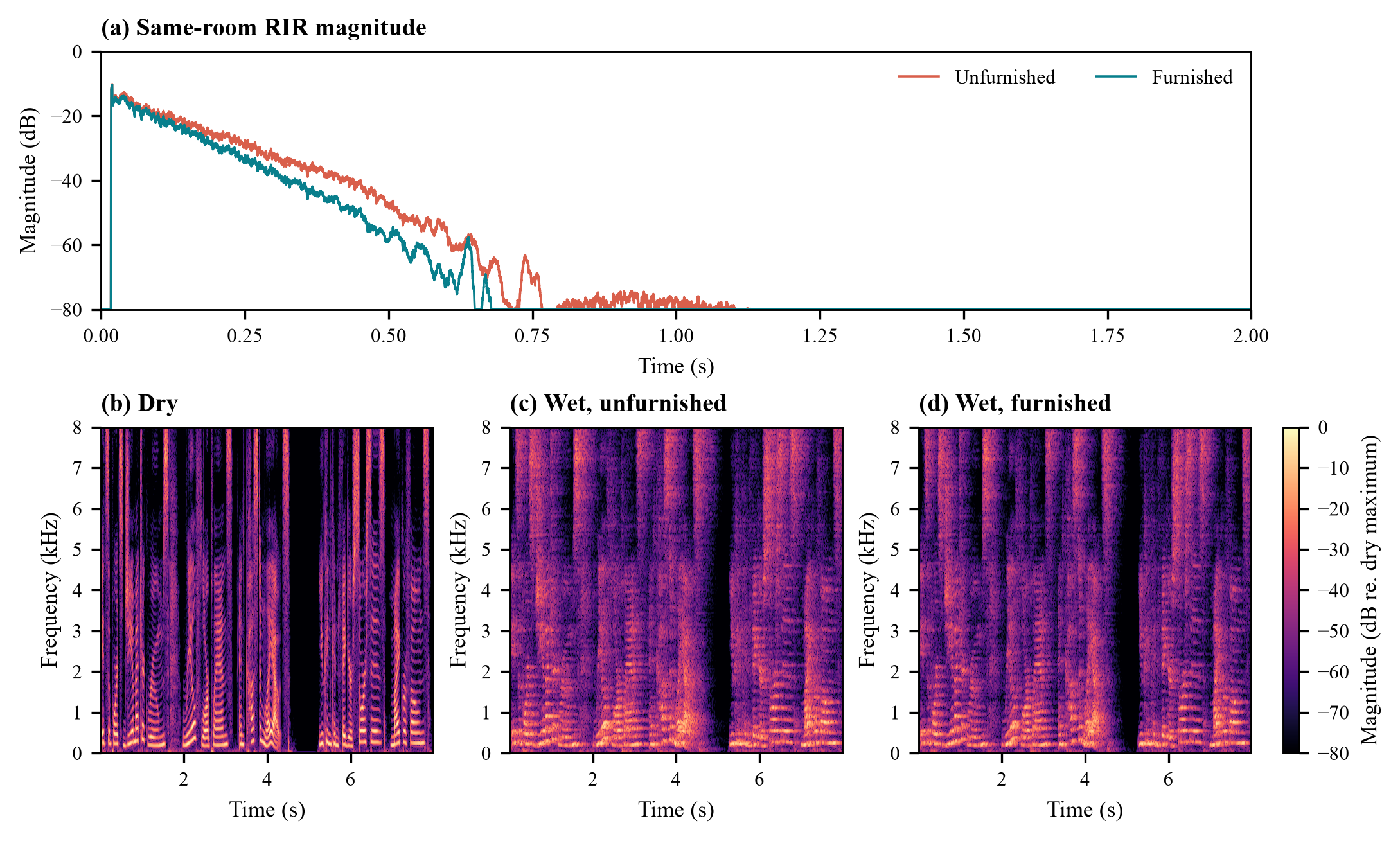}
    \caption{Effect of semantic furniture in a fixed same-room configuration.
    The source and receiver are held at identical positions in the living room
    of FloorPlan~\#5159; geometry, material seed, and solver settings are also
    fixed.  The only change is the addition of 24 automatically placed semantic
    furniture objects.  (a) Unfurnished and furnished RIR magnitude envelopes
    under a common reference.  (b) Dry-speech spectrogram.  (c) Wet speech in
    the unfurnished room.  (d) Wet speech after furniture placement.  All
    spectrograms share the dry-speech maximum as their 0-dB reference.}
    \label{fig:furniture_rir_auralization}
\end{figure}

\subsection{RIR and Auralization Examples}
\label{sec:supp_rir_auralization_examples}

\textbf{Static scene.}
Figure~\ref{fig:static_rir_auralization} shows the complete transformation from
a dry source to a static far-field observation.  The RIR contains a sparse
onset followed by a dense decaying tail.  Relative to the dry spectrogram, the
wet signal exhibits broadened harmonic ridges and residual energy between
speech events, reflecting frequency-dependent attenuation and temporal
smearing by the cross-room propagation path and reverberant field.

\textbf{Dynamic scene.}
The keyframe stack in Fig.~\ref{fig:dynamic_rir_auralization} exposes changes
that a single static RIR cannot represent.  As the receiver approaches the
source, the early response and late-field level evolve continuously.  The wet
spectrogram consequently becomes progressively stronger while preserving
smooth transitions between frames.  This interpolation is identical to the
dynamic auralization procedure used by the Web workbench and avoids abrupt
switching between independently simulated RIRs.

\textbf{Furniture ablation.}
Figure~\ref{fig:furniture_rir_auralization} isolates the acoustic effect of
furniture from source placement and material randomness.  The furnished RIR
decays more rapidly across most of the tail because beds, sofas, rugs, and
other semantic objects introduce additional absorption and scattering area.
The corresponding wet spectrogram contains less persistent inter-event energy
than the unfurnished rendering.  This comparison confirms that semantic
objects participate in acoustic simulation rather than serving only as visual
assets.

\subsection{Interactive Web Workbench}
\label{sec:supp_workbench}

The Web workbench exposes scene construction and acoustic simulation through a
single synchronized interface.  Geometry, FloorPlan, and Custom modes share
the same source, receiver, material, furniture, motion, solver, RIR,
auralization, and API controls.  Editing the scene invalidates the previous
result; running the solver updates the path visualization, metrics, RIR, and
rendered audio from one consistent simulation state.

\begin{figure*}[t]
    \centering
    \includegraphics[width=0.96\linewidth]{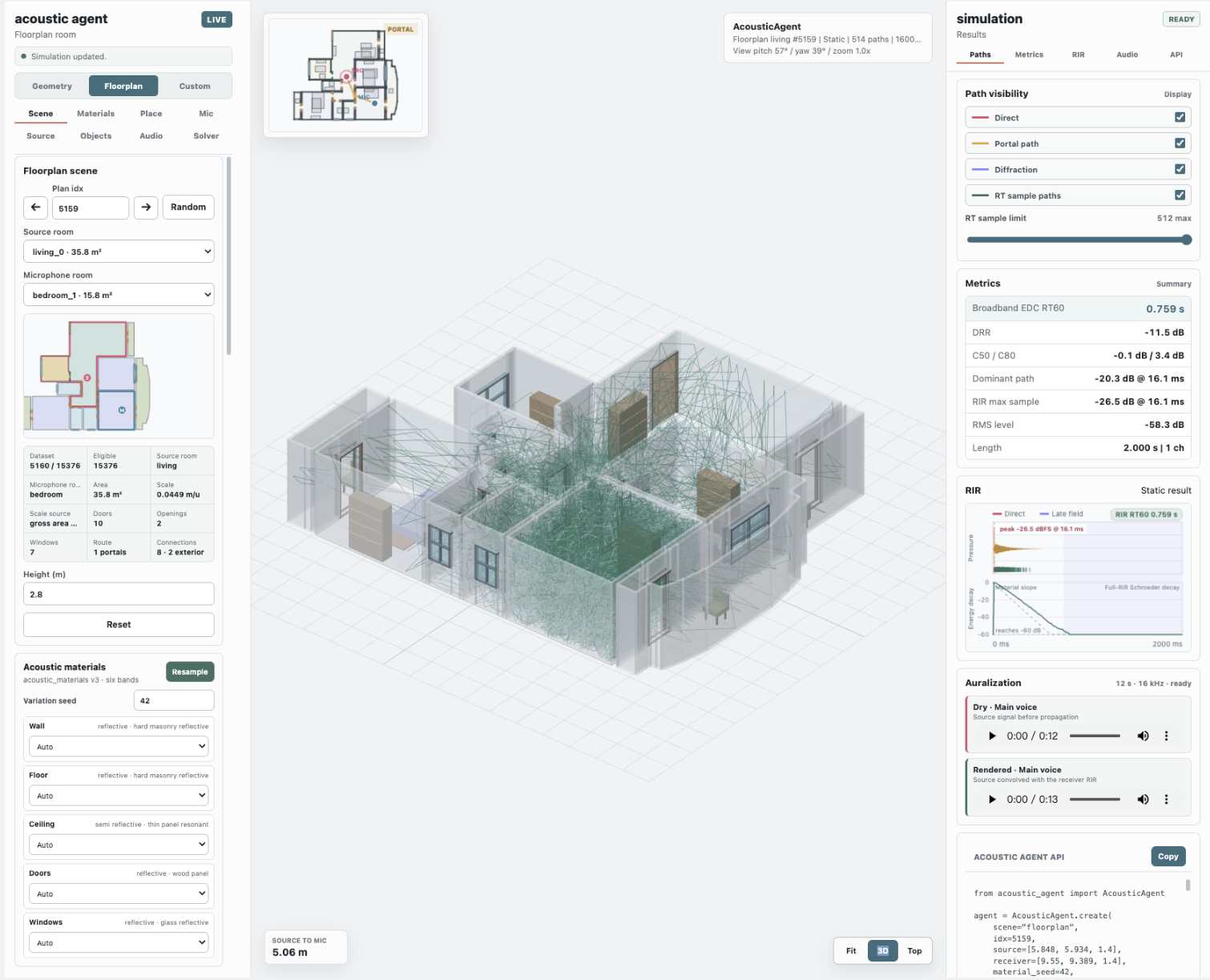}
    \caption{Unified AcousticAgent Web workbench.  The left control rail
    configures the scene, source, microphone, semantic materials, furniture,
    motion, and solver.  The center viewport combines a 2-D floorplan overview
    with an interactive 3-D rendering of the room geometry and traced
    propagation paths.  The right analysis rail exposes path-component
    visibility, acoustic metrics, the rendered RIR, dry/wet auralization, and
    a directly executable Python API example.  All three regions are driven
    by the same scene and simulation result.}
    \label{fig:workbench}
\end{figure*}

Figure~\ref{fig:workbench} shows the FloorPlan mode after a static simulation.
The interface is intended both for qualitative inspection and for checking an
API configuration before batch generation; it does not use a separate or
simplified acoustic solver.

\end{document}